\documentclass[lettersize,journal]{IEEEtran}
\usepackage{amsmath,amsfonts}
\usepackage{algorithmic}
\usepackage{algorithm}
\usepackage{array}
\usepackage[caption=false,font=normalsize,labelfont=sf,textfont=sf]{subfig}
\usepackage{textcomp}
\usepackage{stfloats}
\usepackage{url}
\usepackage{verbatim}
\usepackage{graphicx}
\usepackage{makecell}
\usepackage{cite}
\newtheorem{theorem}{Theorem}
\usepackage{booktabs}
\usepackage{threeparttable}
\usepackage[table]{xcolor}
\usepackage{multirow}
\DeclareRobustCommand{\rev}[1]{#1}
\newcommand{\revcolor}{}

\begin{document}
\bstctlcite{IEEEexample:BSTcontrol}

\title{SELF: A Robust \textbf{S}ingular Value and \textbf{E}igenvalue Approach for \textbf{L}LM \textbf{F}ingerprinting}

\author{Hanxiu Zhang, Qiang Xu,~\IEEEmembership{Senior Member,~IEEE}, Yue Zheng,~\IEEEmembership{Member,~IEEE}
        % <-this % stops a space
\thanks{Hanxiu Zhang and Yue Zheng are with the School of Science and Engineering, The Chinese
University of Hong Kong, Shenzhen, Guangdong 518172, China.
Email: hanxiuzhang@link.cuhk.edu.cn, zhengyue@cuhk.edu.cn.}% <-this % stops a space 

\thanks{Qiang Xu is with the Department of Computer Science and Engineering, The Chinese University of Hong Kong, Hong Kong 999077, China. Email: qxu@cse.cuhk.edu.hk.}

\thanks{Corresponding author: Yue Zheng.}}

\markboth{Journal of \LaTeX\ Class Files,~Vol.~14, No.~8, August~2021}%
{Shell \MakeLowercase{\textit{et al.}}: A Sample Article Using IEEEtran.cls for IEEE Journals}

\IEEEpubid{0000--0000/00\$00.00~\copyright~2021 IEEE}

\maketitle

\begin{abstract}
Safeguarding the Intellectual Property (IP) of Large Language Models (LLMs) against model theft represents a critical challenge in contemporary AI research. Although fingerprinting is essential for establishing forensic evidence for model derivation, existing methods are susceptible to false claim attacks, weight manipulations, or high false positive rates, undermining their utility. To overcome these limitations, we propose SELF, \rev{a novel intrinsic weight-based fingerprinting scheme tailored for robust IP forensics via adversarial model lineage verification}. SELF eliminates dependency on input and inherently resists false claims, achieving robust \rev{provenance detection} through unique, scalable and transformation-invariant fingerprint extraction via singular value and eigenvalue decomposition of LLM attention weights. Experimental results demonstrate that SELF maintains high IP infringement detection accuracy while significantly outperforming state-of-the-art baselines in reducing false positives and showing strong robustness against various downstream modifications, including quantization, pruning, and fine-tuning attacks. Our code is available at \url{https://github.com/HanxiuZhang/SELF_v4}.
\end{abstract}

\begin{IEEEkeywords}
LLM fingerprinting, IP forensics, fingerprinting, singular values, eigenvalues, \rev{model provenance}.
\end{IEEEkeywords}

\section{Introduction}
\label{sec:intro}

% \IEEEPARstart{L}{arge} language models (LLMs) are increasingly being adopted as versatile tools to enhance productivity in various fields such as medical assistance \cite{thirunavukarasu2023large} and code generation \cite{fakhoury2024llm}. \rev{However, as foundation models are widely distributed and increasingly customized through downstream modification, tracking model provenance has emerged as a critical requirement. Determining whether a deployed model is derived from a specific foundational parent is essential not only for enforcing licensing terms and protecting intellectual property (IP) \cite{zheng2024overview}, but also for managing risks, such as tracing the impact of a compromised or vulnerable parent model \cite{nikolic2026model}.} In an era where open-source trends often clash with commercial interests, robust \rev{provenance verification} capabilities are critical for model creators to verify ownership, trace model lineage, and enforce attribution requirements on derivative works.

\IEEEPARstart{L}{arge} language models (LLMs) are increasingly being adopted as versatile tools to enhance productivity in various fields such as medical assistance \cite{thirunavukarasu2023large} and code generation \cite{fakhoury2024llm}. However, developing a functional LLM requires substantial investment spanning high-quality datasets, significant computational resources, and specialized human expertise. These high costs make LLMs attractive targets for model theft and unfair competition. Consequently, Intellectual Property (IP) forensics has emerged as an essential requirement for LLM protection\cite{zheng2024overview}, \rev{with reliable model provenance verification serving as its fundamental core}. In an era where open-source trends often clash with commercial interests, robust \rev{provenance verification} capabilities are critical for model creators to verify ownership, trace model lineage, and enforce attribution requirements on derivative works. 

To \rev{trace model provenance}, current methods primarily fall into two categories: watermarking and fingerprinting. Watermarking approaches embed identifiable features (watermarks) invasively into target models while trying to preserve their original functionality \cite{zhang2024remark,xu2024instructional}. In contrast, fingerprinting methods extract unique model identifiers without modifying the model, either by analyzing the model’s input-output behavioral patterns \cite{gubri2024trap} (i.e., behavior fingerprinting) or structural information (i.e., structural fingerprinting) such as weight distributions \cite{zeng2024huref}, intermediate representations \cite{zhang2025reef}, or gradient profiles \cite{wu2025gradientbasedmodelfingerprintingllm}. \rev{For practical post-deployment provenance testing}, fingerprinting offers a decisive advantage. Unlike watermarks that must be embedded prior to release, fingerprinting schemes enable post-hoc forensics, allowing owners to verify the \rev{provenance} of models already in circulation without the need of retraining or risking potential performance degradation \cite{zheng2022dnn}.

\begin{figure}[t]
\centering
\includegraphics[width=\linewidth]{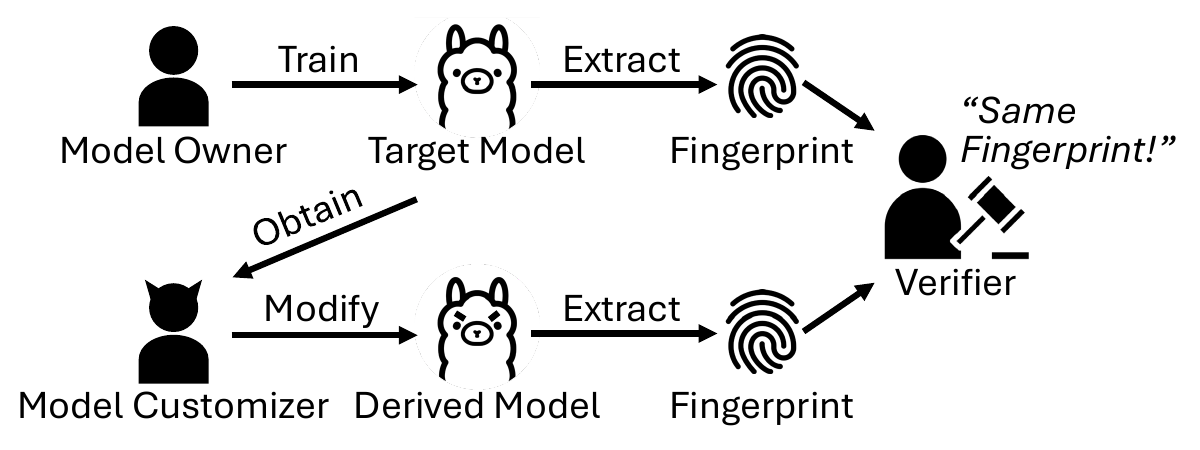}
\makeatletter
\renewcommand{\fnum@figure}{\rev{\figurename~\thefigure}}
\makeatother
    \caption{\rev{Model provenance verification} pipeline using SELF. 
    After training the target model, the owner can extract the model fingerprint for \rev{lineage tracking}. Even if the \rev{derived} model has been modified to evade detection, the owner can still \rev{verify provenance} by extracting and comparing the models' fingerprints. A high fingerprint similarity provides evidence of \rev{derivation from the foundational target model}.
    }
    \label{fig:pipeline}
\end{figure}

\IEEEpubidadjcol

Despite these advantages, existing fingerprinting methods face critical limitations. Behavior-based techniques are vulnerable to false claim attacks \cite{liu2024false}, wherein malicious actors can falsely claim the ownership of independently trained models by crafting (transferable) adversarial samples. Although Shao et al. \cite{shao2025fitprintfalseclaimresistantmodelownership} propose mitigating the attack by constructing fingerprints using targeted adversarial examples, the risk persists as such adversarial examples can still be transferable albeit with greater difficulty. Structural approaches analyze model internal parameters but often lack robustness against weight manipulations or suffer from a high false positive rate (FPR). 
We demonstrate in paragraph a) of Section \ref{sec: false_claim} that even the state-of-the-art (SOTA) work, REEF \cite{zhang2025reef}, exhibits a high FPR, leading to the erroneous flagging of independent models as \rev{derived} models. 
%Furthermore, for structural schemes like HuRef \cite{zeng2024huref} where the input is required to actively participate in fingerprint computation, we further extend the scope of false claim attack as a malicious accuser can manipulate \rev{provenance verification results} through carefully crafted input. 
Beyond such false positives, input-dependent structural fingerprinting methods such as HuRef \cite{zeng2024huref} are vulnerable to direct input manipulation: a malicious accuser can optimize input tokens to artificially increase the fingerprint similarity between unrelated models, thereby manipulating provenance verification results without modifying either model’s parameters.
Under this definition, we conduct false claim attack on the HuRef scheme and successfully manipulate the similarity score output (Detailed experiments are provided in paragraph b) of Section \ref{sec: false_claim}). 

To address these practical deployment issues, we propose a structural fingerprinting method named SELF, which purely depends on the model weights to \rev{verify provenance}. Fig. \ref{fig:pipeline} describes SELF's pipeline. The core principle is that fingerprints extracted from the target model and its \rev{derivatives} will exhibit high similarity, enabling reliable \rev{provenance detection}. During the fingerprint extraction stage, we address potential model weight tampering caused by transformation attacks (e.g., permutation and linear-mapping \cite{zeng2024huref}) through identifying invariant attributes. Specifically, we first compute singular value and eigenvalue invariant matrices from the weight matrices, then derive their corresponding invariant singular values and eigenvalues to construct the model fingerprint. Leveraging the inherent properties of these values, the fingerprint remains both invariant against transformation attacks and robust to downstream modifications such as quantization and fine-tuning. 

Our contributions can be summarized as follows:
\begin{enumerate}
\item We present SELF, a weight-based fingerprinting method for \rev{verifying LLM provenance}. This weight-exclusive methodology eliminates dependency on input samples, thereby preventing false claim of \rev{provenance} caused by manipulating adversarial inputs. 
\item We propose robust fingerprint extraction against transformation attacks. By deriving fingerprints from singular values and eigenvalues of matrices constructed from model weights, our approach leverages their fundamental mathematical properties to ensure uniqueness, scalability, and robustness. 
\item We identify and empirically validate critical security flaws in SOTA structural fingerprinting schemes. Specifically, we demonstrate that input-dependent schemes like REEF \cite{zhang2025reef} suffers from prohibitively high FPR, while HuRef \cite{zeng2024huref} are vulnerable to input-manipulation false claim attacks (Section \ref{sec: false_claim}). 
\item Our extensive evaluation on model families including Qwen2.5-7B and Llama2-7B demonstrates that SELF achieves both accurate and robust \rev{provenance verification}. The method maintains its discriminative capability even when subjected to various model modifications, including quantization, pruning, and fine-tuning attacks, significantly outperforms SOTA structural baselines by eliminating false positives and preventing false claim attacks.

\end{enumerate}

The rest of the paper is organized as follows. Section \ref{sec:preliminary} provides the preliminaries, covering the attention mechanism in LLMs and transformation attacks designed to evade detection. Section \ref{sec:sv_ev} investigates singular values and eigenvalues as invariant fingerprints robust to transformation attacks. Section \ref{sec:methodology} details the proposed methodology, including the threat model, fingerprint extraction pipeline, and similarity computation process. Section \ref{sec:experiment} evaluates the effectiveness and robustness of our method compared to existing baselines. The existing \rev{model provenance verification} approaches are reviewed in Section \ref{app:related_work}. Finally, Section \ref{sec:conclusion} concludes the paper \rev{and discusses limitations and future research directions}.

\section{Preliminaries}
\label{sec:preliminary}

\subsection{Attention Mechanism}
\label{sec:attention}

LLMs widely adopt the Transformer architecture \cite{vaswani2017attention}, with the attention mechanism serving as its core component. The computation of this mechanism can be formulated as:

\begin{equation}
    H_{\text{out}} = \mathrm{softmax}\left(\frac{H_{\text{in}} W_Q (H_{\text{in}} W_K)^\top}{\sqrt{d}}\right)(H_{\text{in}} W_V) W_O
    \label {eq:attn}
\end{equation}

where \(H_{\text{in}}\), \(H_{\text{out}} \in \mathbb{R}^{n \times d_{\text{model}}} \) denote the input hidden representations and the self-attention output, respectively,  \(W_Q, W_K, W_V \in \mathbb{R}^{d_{\text{model}} \times d}\) represent the learnable weight matrices for the Query, Key, and Value matrices, respectively, while \(W_O \in \mathbb{R}^{d \times d_{\text{model}}}\) denotes the output weight matrix. Here, \(n\), \(d\) and \(d_{\text{model}}\) correspond to sequence length, projection dimension, and embedding dimension, respectively. 

\subsection{Transformation Attacks}
\label{sec:trans_atk}
Model weights are a direct consequence of the specific learning process the model goes through and the data it was trained on, therefore can serve as unique fingerprints for model similarity comparison. However, weight-based fingerprint verification may  be evaded by the following transformation attacks, which aim to preserve the model’s functional behavior but significantly modify the weights \cite{zeng2024huref}:

\paragraph{Permutation Attack}
This attack rearranges the weights by permuting the weight matrix. Let \(P \in \mathbb{R}^{d_{\text{model}} \times d_{\text{model}}}\) denote an arbitrary permutation matrix. This permutation transformation needs to be applied to the input of the model and corresponding weights: 
\begin{equation}
\begin{split}
    \hat{H}_{\text{in}} &= H_{\text{in}} P^T, \\
    \hat{W}_Q &= P W_Q, \quad 
    \hat{W}_K = P W_K, \quad \\
    \hat{W}_V &= P W_V, \quad 
    \hat{W}_O = W_O P^T.
\end{split}
\label{eq: per_attack}
\end{equation}
As a result, \(\hat{H}_{\text{out}} = H_\text{out} P^T\). The permutation effect propagates through subsequent layers, making each layer's input and output be permuted with \(P^T\). The original model function can be preserved by applying the same permutation \(P\) to the final output. 

\paragraph{Linear Mapping Attack}
Alternatively, a pair of linear transformations can be applied to the attention weights:
\begin{equation}
\begin{split}
    \hat{W}_Q &= W_Q C_1, \quad \hat{W}_K = W_K C_1^{-T} , \\ \quad \hat{W}_V &= W_V C_2, \quad \hat{W}_O = C_2^{-1} W_O.
\end{split}
\label{eq: lin_map_attack}
\end{equation}
where \(C_1, C_2 \in \mathbb{R}^{d \times d}\) are arbitrary invertible matrices and $T$ denotes the transpose. This linear transformation preserves the attention scores, e.g., 
\begin{equation}
\begin{split}
(H_{\text{in}} \hat{W}_Q)(H_{\text{in}}\hat{W}_K)^T &= H_{in}W_QC_1C_1^{-1} W_K^TH_{in}^T \\
& = (H_{\text{in}} W_Q)(H_{\text{in}} W_K)^T, \\
(H_{\text{in}}\hat{W}_V)\hat{W}_O &= (H_{\text{in}}W_VC_2)C_2^{-1}W_O \\
&=(H_{\text{in}}W_V)W_O,
\end{split}
\end{equation}
leaving the output \(H_\text{out}\) unchanged.

The two attacks highlight a fundamental vulnerability: the model output remains unaffected, while the model weights are altered in ways that invert similarity metrics computed purely on weight comparison.

\section{Exploring Singular Values and Eigenvalues as Invariant Fingerprints}
\label{sec:sv_ev}

Singular values and eigenvalues are fundamental matrix attributes that capture intrinsic properties of a matrix. \rev{This section introduces the definitions of the two features and their transformation invariance, providing the mathematical rationale for our invariant matrix design in Sec.~\ref{sec:fingerprint}.}

% \rev{This section reviews motivating linear-algebraic facts about singular values and eigenvalues to provide the mathematical rationale for our invariant matrix design in Sec.~\ref{sec:fingerprint}. Specifically, we introduce the computation of these quantities and discuss the invariance properties relevant to transformation attacks.}

\subsection{Singular Values}
Singular values are defined for matrices with any size and describe how much the matrix stretches space along principal mutually orthogonal directions. Any complex matrix \( M \in \mathbb{C}^{m \times n} \) admits a singular value decomposition (SVD):
\begin{equation}
    M = U \Sigma V^*,
\end{equation}
where \( U \in \mathbb{C}^{m \times m} \) and \( V \in \mathbb{C}^{n \times n} \) are unitary matrices. The matrix \( \Sigma \in \mathbb{R}^{m \times n} \) is a rectangular diagonal matrix with non-negative real numbers \( \sigma_1 \geq \sigma_2 \geq \cdots \geq 0 \) on the diagonal. These values \( \sigma_i = \Sigma_{ii} \) for \( i = 1,2,\dots,\min\{m,n\} \) are uniquely determined by \( M \) and are referred to as the \emph{singular values} of \( M \).
\paragraph{Singular values remain unchanged under row, column permutation or any combination of both} Specifically, let \( P_1 \in \mathbb{R}^{m \times m} \), \( P_2 \in \mathbb{R}^{n \times n} \) be the row and column permutation matrices, respectively, and $\sigma_i(\cdot)$   denote the $i_{th}$ singular value of the corresponding matrix. Since permutation matrices are orthogonal, we have: 
\begin{equation}
\sigma_i(M)=\sigma_i(P_1 M) = \sigma_i(M P_2) = \sigma_i(P_1MP_2). 
\end{equation}
This property \cite{franklin2000matrix} enables the use of singular values as robust descriptors for model weight matrices, providing resistance to the permutation attacks defined in Equation \eqref{eq: per_attack} through the construction of singular value invariant matrix in Section \ref{sec:fingerprint}.
\paragraph{Singular values exhibit stability under small perturbations} Let $\Delta M$  denote the perturbations added to matrix $M$, according to Weyl's inequality \cite{franklin2000matrix}: 
\begin{equation}
    \left| \sigma_i(M + \Delta M) - \sigma_i(M) \right| \leq \| \Delta M \|_2,
\label{eq:weyl}
\end{equation}
where \( \| \cdot \|_2 \) denotes the spectral norm. This implies that the singular values of a matrix remain approximately unchanged as long as the perturbation magnitude \( \| \Delta M \|_2 \) is small.

\subsection{Eigenvalues}
Eigenvalues are defined only for square matrices   and describe the scaling factors associated with specific invariant directions (eigenvectors). For a square matrix \( N \in \mathbb{C}^{n \times n} \), the eigenvalues \( \lambda_1, \ldots, \lambda_n \) are defined as the roots of its characteristic polynomial, which are obtained by solving:
\begin{equation}
    \det(N - \lambda I)=0,
\end{equation}
where $\det$  denotes determinant and $I$ is the identity matrix. If \( N \) is diagonalizable, eigenvalues can be more efficiently and stably computed via the following eigenvalue decomposition (EVD):
\begin{equation}
    N = Q \Lambda Q^{-1},
\end{equation}
where \( Q \in \mathbb{C}^{n \times n} \) consists of the linearly independent eigenvectors of \( N \), and \( \Lambda \in \mathbb{C}^{n \times n} \) is a diagonal matrix whose diagonal entries \( \Lambda_{ii} = \lambda_i \) for \( i = 1,2,\dots,n \) are the eigenvalues of \( N \).  
\paragraph{Eigenvalues are invariant under similarity transformations} Let \( C \) be an invertible matrix, the operation of transforming a matrix $N$ to \( \hat{N} = C N C^{-1} \) is called a similarity transformation. The eigenvalues of $\hat{N}$ can be obtained by solving the roots of 
\begin{equation}
\begin{aligned} 
    & \det(\hat{N}-\lambda I) = \det(C N C^{-1} -\lambda I) \\
    & = \det(C N C^{-1} - \lambda I(CC ^{-1})) =\det(C N C^{-1} -C(\lambda I)C ^{-1}) \\
    & = \det(C(N-\lambda I)C^{-1})  = \det(C) \det(N-\lambda I)\det(C^{-1}) \\
    & = \det(C) \det(N-\lambda I)\frac{1}{\det(C)} = \det(N-\lambda I).\end{aligned}
\end{equation}
Therefore, \( \hat{N} \) and \( N \) share the same eigenvalues. This property forms the theoretical foundation for defending against the linear mapping attack   described in Equation \eqref{eq: lin_map_attack} through the construction of eigenvalue invariant matrix in Section \ref{sec:fingerprint}.
\rev{
\paragraph{Eigenvalues are robust against small perturbations} When $N$ is diagonalizable as $N = Q \Lambda Q^{-1}$, the columns of $Q$ are the right eigenvectors, while the rows of $Q^{-1}$ correspond to the conjugate transposes of the left eigenvectors. For a simple eigenvalue $\lambda_i$ with normalized right eigenvector $x_i$ and left eigenvector $y_i$ ($\|x_i\|_2 = \|y_i\|_2 = 1$), first-order perturbation theory \cite{greenbaum2020first} states that for a perturbed matrix $N + \Delta N$, the shift in the eigenvalue satisfies:

\begin{equation}
|\lambda_i(N + \Delta N) - \lambda_i(N)| \le \frac{\|\Delta N\|_2}{|y_i^* x_i|} + O(\|\Delta N\|_2^2) 
\label{eq:fop}
\end{equation}
where the term $1/|y_i^* x_i|$ is the condition number of the eigenvalue $\lambda_i$. This condition number explicitly characterizes the local sensitivity of $\lambda_i$ to matrix perturbations. }

In the context of model fingerprinting, both singular values and eigenvalues can serve as transformation-invariant descriptors, providing robustness against adversarial manipulations of model weights. 

\begin{figure*}[ht!]
    \centering
    \includegraphics[width=0.95\textwidth]{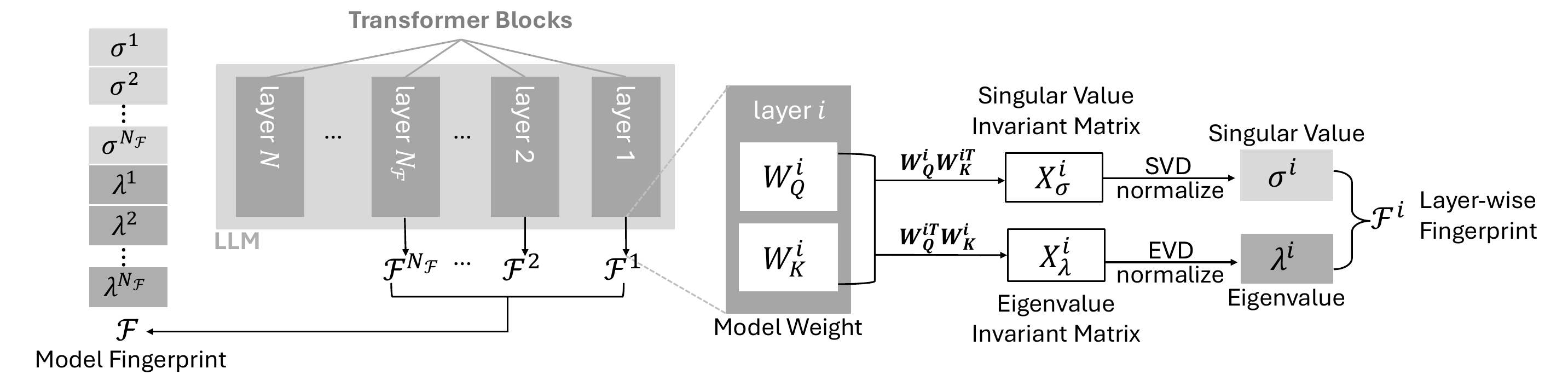}  
    \caption{Fingerprint extraction via matrix decomposition. For a given model, we extract its fingerprint using the first \(N_\mathcal{F}\) Transformer block layers. Specifically, we first compute the fingerprint of each individual layer \(\mathcal{F}^i\) and then aggregate the \(N_\mathcal{F}\) layer-wise fingerprints to form the overall model fingerprint \(\mathcal{F}\). For each layer, we first compute the singular value invariant matrix \(X_{\sigma}^i\) and the eigenvalue invariant matrix \(X_{\lambda}^i\) from the attention weights \(W_Q^i\) and \(W_K^i\). Then, we extract the normalized singular value vector \(\sigma^i\)and eigenvalue vector \(\lambda^i\), which together form the fingerprint of layer \(i\). 
}
    \label{fig:methodology-1}
\end{figure*}

\section{Methodology}
\label{sec:methodology}

Our method consists of two components: 1) Fingerprint Extraction. This component analyzes the model weights to extract unique and scalable fingerprints by leveraging the properties of singular values and eigenvalues. 2) Similarity Computation. This component detects fingerprint similarity through distance metric. The following sections introduce the threat model and these essential components. 

\subsection{Threat Model}

\rev{
We specifically align our threat model with the model provenance verification scenario. Unlike black-box behavioral fingerprinting typically used for initial suspicion, this scenario represents a distinct forensic phase where definitive, legally admissible proof of derivation is mandated. Our threat model involves three main roles: the model owner (the original developer or an authorized distributor), the model customizer (the potential adversary), and the verifier (a trusted third party or forensic auditor). The model owner holds the ownership or licensing rights to the \textbf{target model} (the model to be protected) and retains white-box access to its parameters. The model customizer obtains the target model (e.g., via open-source release, insider leakage, or commercial licensing) and may derive a new model from it through fine-tuning, pruning, quantization, or weight transformations while preserving its functionality. The customizer may also apply false-claim attacks to obscure the model's provenance. Once the model owner flags a \textbf{suspect model} (suspicious to be pirated), he/she could request model provenance verification procedures by the verifier.}

The mechanism must distinguish between: \textbf{related models} (derived from the target, even if modified) and \textbf{unrelated models} (independently trained), while operating without knowledge of the specific attack employed. We posit that the verifier temporarily obtains white-box access to both the target and suspect models only during the investigation, and its objective is to reliably verify model provenance despite potential adversarial modifications. We argue that this white-box assumption is not a limitation but a necessity for forensic verification. \rev{Such authorized access enables the verifier to examine internal model parameters as direct evidence, making SELF highly applicable in scenarios such as model auditing, lineage analysis, and forensic investigation.}
% The verification mechanism must distinguish between \textbf{related models}, which are derived from the target model and may have undergone substantial modifications, and \textbf{unrelated models}, which are independently developed or trained. This distinction must be made without prior knowledge of the type or extent of the applied modifications. We assume that the verifier is granted temporary white-box access to both the target and suspect models during the verification process, with the objective of reliably determining their provenance relationship despite possible benign or adversarial transformations. This white-box assumption defines the intended verification setting and enables the verifier to examine internal model parameters as direct evidence of derivation. Such access may be provided to a trusted third party in applications such as model auditing, lineage analysis, or forensic investigation.
% }

%Our method aims to provide accurate, robust, and efficient infringement detection for IP forensic scenarios. 

\subsection{Fingerprint Extraction via Matrix Decomposition}
\label{sec:fingerprint}
\subsubsection{Transformation Invariant Matrices}
To extract robust fingerprint, we focus on invariant features derived from the core component of the LLM, i.e., the attention blocks.
% Attention blocks are selected for their robust discriminability, high dimensionality, and architectural university. 
Compared to other components, the advantage of the attention block lies in (1) High-dimensionality: Attention matrices encode rich, high-dimensional features (e.g.,$W_Q$, $W_K$ account for $>$15\% of Llama2-7B's parameters), while LayerNorm has a much smaller parameter size (e.g., 0.1\% of Llama2-7B's total parameters) and outputs vectors, which limit its utility for SVD- and EVD-based fingerprint extraction. 
(2) Architectural generality: Attention blocks are universal across transformer variants, while other blocks such as MLPs (e.g., in Mixture of Experts models \cite{shazeer2017outrageously}) exhibit structural variability that could undermine consistency. 
\rev{
Within attention blocks, we specifically select the query and key projection matrices, $W_Q$ and $W_K$, based on their distinct functional roles~\cite{elhage2021mathematical}. Specifically, $W_Q$ and $W_K$ determine attention routing, whereas $W_V$ and $W_O$ determine the content written to the residual stream, thereby influencing subsequent computations and output logits. This functional distinction motivates deriving the invariants from $W_Q$ and $W_K$, as they tend to be comparatively stable when downstream adaptation primarily modifies factual knowledge or stylistic behavior while largely preserving attention routing.
}

Let \( W_Q \in \mathbb{R}^{d_{\text{model}} \times d} \) and \( W_K \in \mathbb{R}^{d_{\text{model}} \times d} \) denote the query and key projection weight matrices in a Transformer attention block, where \( d_{\text{model}} \) is the embedding dimension and \( d \) is the projection dimension. As described in Section \ref{sec:trans_atk}, these matrices may be subjected to the permutation and linear mapping attacks of the following form:

\begin{equation}
    \hat{W}_Q = P W_Q C_1, \quad \hat{W}_K = P W_K C_1^{-T}.
    \label{eq:attack_form}
\end{equation}
where \( P \in \mathbb{R}^{d_{\text{model}} \times d_{\text{model}}} \) is a permutation matrix, and \( C \in \mathbb{R}^{d \times d} \) are invertible matrices. To counter these attacks, we introduce two transformation-invariant matrices for robust fingerprinting.

\textbf{Singular Value Invariant Matrix:}
We first define matrix \(X_\sigma\) that preserves singular values under the transformation attacks in Equation \eqref{eq:attack_form}:
\begin{equation}
\begin{split}
    &X_\sigma = W_Q W_K^T \in \mathbb{R}^{d_{\text{model}} \times d_{\text{model}}}.
    % &Y_\sigma = W_V W_O \in \mathbb{R}^{d_{\text{model}} \times d_{\text{model}}}
\end{split}
    \label{eq:sigma_matrix}
\end{equation}

\begin{theorem}[Singular Value Invariance]
\label{sv_invariance}
Under the transformation attack described in Equation \eqref{eq:attack_form}, the matrix \(\hat{X}_\sigma = \hat{W}_Q \hat{W}_K^T\) satisfies:
\begin{equation}
\begin{split}
        \hat{X}_\sigma &= (P W_Q C_1)(P W_K C_1^{-T})^T \\
                       &= P W_Q C_1 C_1^{-1} W_K^T P^T \\
                       &= P W_Q W_K^T P^T = P X_\sigma P^T.
        % &\hat{Y}_\sigma = (P W_V C_2)(C_2^{-1} W_O P^T) = P W_V C_2 C_2^{-1} W_O P^T = P W_V W_O P^T = P Y_\sigma P^T
\end{split}
\end{equation}
Since permutation matrices are orthogonal, \(X_\sigma\) and \(\hat{X}_\sigma\) are orthogonally equivalent and consequently share the same singular values.
\end{theorem}

\textbf{Eigenvalue Invariant Matrix:}
Next, we define matrix \(X_\lambda\) that preserves eigenvalues under the transformation attack in Equation \eqref{eq:attack_form}:

\begin{equation}
\begin{split}
    &X_\lambda = W_Q^T W_K \in \mathbb{R}^{d \times d}.
    % &Y_\lambda = W_O W_V \in \mathbb{R}^{d \times d}
\end{split}
    \label{eq:lambda_matrix}
\end{equation}

\begin{theorem}[Eigenvalue Invariance]
\label{ev_invariance}
Under the transformation attack described in Equation \eqref{eq:attack_form}, the matrix \(\hat{X}_\lambda = \hat{W}_Q^T \hat{W}_K\) satisfies:

\begin{equation}
    \begin{split}
        \hat{X}_\lambda &= (P W_Q C_1)^T(P W_K C_1^{-T}) \\
                        &= C_1^T W_Q^T P^T P W_K C_1^{-T} \\
                        &= C_1^T W_Q^T W_K C_1^{-T} = C_1^T X_\lambda C_1^{-T}.
        % & \hat{Y}_\lambda = (C_2^{-1} W_O P^T)(P W_V C_2) = C_2^{-1} W_O P^T P W_V C_2 = C_2^{-1} W_O W_V C_2 = C_2^{-1} Y_\lambda C_2
    \end{split}
\end{equation}

This establishes a similarity transformation between \(X_\lambda\) and \(\hat{X}_\lambda\), guaranteeing each matrix pair share identical eigenvalues.
\end{theorem}

\subsubsection{Proposed Fingerprints}
Fig. \ref{fig:methodology-1} shows the fingerprint extraction pipeline. Given an LLM, we extract its fingerprint by analyzing the first \(N_\mathcal{F}\) Transformer block layers. For each layer \(i\), we compute a layer-wise fingerprint \(\mathcal{F}^i\). The overall fingerprint \(\mathcal{F}\)  of the model can be obtained by aggregating all \(N_\mathcal{F}\) layer-wise fingerprints.

\textbf{Fingerprint for a Single Layer:}  
 Let \(W_Q^i\) and \(W_K^i\) denote the query and key matrices of the $i_{th}$ Transformer block layer. We compute the singular value invariant matrix \(X_\sigma^i\) as defined in Equation \eqref{eq:sigma_matrix} and the eigenvalue invariant matrix \(X_\lambda^i\) as defined in Equation \eqref{eq:lambda_matrix}. From the two matrices, we extract the following features sorted in descending order:
\begin{enumerate}
\item the top \(h\) singular values of \(X_\sigma^i\), denoted by vector \(\tilde{\sigma}^i \in \mathbb{R}^{h}\), 
\item the top \(h\) eigenvalues (by magnitude) of \(X_\lambda^i\), denoted by  vector \(\tilde{\lambda}^i \in \mathbb{R}^{h}\), 
% \item the top \(h\) singular values of \(Y_\sigma^i\), denoted by vector \(\tilde{\sigma}^i_{VO} \in \mathbb{R}^{h}\), 
% \item the top \(h\) eigenvalues (by magnitude) of \(Y_\lambda^i\), denoted by  vector \(\tilde{\lambda}^i_{VO} \in \mathbb{R}^{h}\), 
\end{enumerate}
then each vector is normalized to unit \(L_2\) norm, e.g.,
\(
    \sigma^i = \frac{\tilde{\sigma}^i}{\|\tilde{\sigma}^i\|_2}
    % \lambda^i_{QK} = \frac{\tilde{\lambda}^i_{QK}}{\|\tilde{\lambda}^i_{QK}\|_2},
    % \sigma^i_{VO} = \frac{\tilde{\sigma}^i_{VO}}{\|\tilde{\sigma}^i_{VO}\|_2}, 
    % \lambda^i_{VO} = \frac{\tilde{\lambda}^i_{VO}}{\|\tilde{\lambda}^i_{VO}\|_2}
\), 
and the fingerprint for layer \(i\) is then defined with the normalized vectors as 
\(
    \mathcal{F}^i = [\sigma^i,\lambda^i] \in \mathbb{R}^{2 \times h}
\).

\textbf{Fingerprint for the Whole Model:}  
As studied by \cite{zheng2022dnn},  lower-layer (near to the input) weights tend to be more robust against fine-tuning or task-specific adaptations. Therefore, we extract fingerprints from the first \(N_{\mathcal{F}}\) Transformer layers. The full model fingerprint is obtained by stacking all singular and eigenvalue vectors in the following order:

\begin{equation}
\begin{split}
    \mathcal{F} = [
    &\sigma^1,
    \sigma^2,
    \dots,
    \sigma^{N_{\mathcal{F}}},
    \lambda^1,
    \lambda^2,
    \dots,
    \lambda^{N_{\mathcal{F}}}
    % &\sigma^1_{VO},
    % \sigma^2_{VO},
    % \dots,
    % \sigma^{N_{\mathcal{F}}}_{VO},
    % \lambda^1_{VO},
    % \lambda^2_{VO},
    % \dots,
    % \lambda^{N_{\mathcal{F}}}_{VO}
    ]^T \in \mathbb{R}^{2N_{\mathcal{F}} \times h}
\end{split}
\end{equation}
This fingerprint matrix \(\mathcal{F}\) captures intrinsic structural information from multiple attention layers, while remaining robust to the weight transformation attacks.

\subsection{Fingerprint Similarity Detection via $L_1$ Distance}

To determine whether a suspect model $S$ is a derived variant of the target model $T$, we employ the $L_1$ distance to quantify fingerprint similarity. \rev{This choice is intrinsically aligned with our theoretical perturbation analysis in Eq. \ref{eq:weyl} and \ref{eq:fop}, which bounds the change of each component in terms of its absolute deviation.} Let $\mathcal{F}_{T}$ and $\mathcal{F}_{S}$ denote the fingerprints of the target and suspect models, respectively. We calculate the distance:
\begin{equation}
    dist = \|\mathcal{F}_{T} - \mathcal{F}_{S}\|_1
\end{equation}
If $dist < \tau$, where $\tau>0$ is a predefined threshold, $S$ is identified as a potential derivative of $T$. Conversely, if the distance exceeds the threshold, $S$ is considered an unrelated model.

\section{Experiments}
\label{sec:experiment}

This section evaluates SELF in terms of effectiveness and robustness, and provides a comprehensive comparison with SOTA structural fingerprinting methods.  Unless otherwise specified, we extract fingerprints from the first \(N_\mathcal{F} = 8\) layers, using the top \(h = 256\) singular values and eigenvalues, resulting in fingerprint size of \(16\times 256\). Systematic ablation studies on the sensitivity of SELF's performance to variations in layer selection, weight matrix choices, $h$, and $N_\mathcal{F}$ are also evaluated. Finally, we demonstrate a concrete false claim attack on HuRef, revealing a critical vulnerability inherent in current input-dependent fingerprinting schemes. 

\subsection{Effectiveness Verification}
\label{sec:eff_ve}

We use Qwen2.5-7B and Llama2-7B as target models to evaluate the effectiveness of SELF, and their publicly available offsprings are used as related models. For unrelated models, we select 10 independent open-source models with diverse architectures and parameter sizes ranging from 800M to 7B.
As shown in Table \ref{tab:sim}, the $L_1$ distances of related models are \textbf{$<$ 21} while those for unrelated models are \textbf{$>$ 34}, demonstrating that our method can effectively discriminate between related and unrelated models\footnote{\rev{Based on this margin and the systematic sensitivity analysis detailed later in Sec.~\ref{sec: lb_ablation}, we formally set the decision threshold $\tau$ to 24.}}. \rev{Notably, the upper bound of the related models' distance is primarily driven by domain-specialized variants (e.g., Qwen2.5-Coder-7B at 20.9766). The fact that even industrial-scale continued learning on massive alternative-domain data fails to push the fingerprint distance into the unrelated regime demonstrates SELF's resilience against extreme domain adaptation.}

\begin{table*}[h!]
\centering
\belowrulesep=0pt
\aboverulesep=0pt
\renewcommand{\arraystretch}{1.25}
\caption{Fingerprint $L_1$ Distance (Target Models: Qwen2.5-7B and Llama2-7B)}
\begin{tabular*}{\linewidth}{@{\extracolsep{\fill}}lclc|lclc}
%\begin{tabular}{@{} l c @{\hspace{1.2em}} l c  @{\hspace{1.2em}} |
                % l c @{\hspace{1.2em}} l c @{}}
\toprule
\multicolumn{2}{c}{\textbf{Qwen2.5-7B Unrelated}} &  
\multicolumn{2}{c|}{\textbf{Qwen2.5-7B Related}} &  
\multicolumn{2}{c}{\textbf{Llama2-7B Unrelated}} &  
\multicolumn{2}{c}{\textbf{Llama2-7B Related}} \\
\textbf{Model} & \textbf{Distance} & \textbf{Model} & \textbf{Distance} & \textbf{Model} & \textbf{Distance} & \textbf{Model} & \textbf{Distance} \\
\midrule
Mistral-7B-V0.3 & 44.9266 & \textbf{Fine-tuned Variants} &             & Mistral-7B-V0.3 & 34.5286   & \textbf{Fine-tuned Variants} &  \\
Llama2-7B       & 57.4234 & Qwen2.5-7B-Instruct          & 0.6330     & Qwen1.5-7B      & 44.3750   & Llama-2-7B-Chat              & 0.8223 \\
Baichuan2-7B    & 70.0009 & Qwen2.5-Math-7B              & 13.2143    & Baichuan2-7B    & 48.5008   & CodeLlama-7B                 & 12.6612 \\
InternLM2.5-7B  & 55.7048 & Qwen2.5-Coder-7B             & 20.9766    & InternLM2.5-7B  & 65.0354   & Llemma-7B                    & 12.7050 \\
GPT2-Large      & 66.6810 & TableGPT2-7B                 & 1.3130     & GPT2-Large      & 49.9396   & \textbf{Pruned Variants}     &  \\
Cerebras-GPT-1.3B & 88.0211 & Qwen2.5-7B-Medicine        & 0.6833     & Cerebras-GPT-1.3B & 67.3757 & Sheared-Llama-2.7B           & 9.8122 \\
ChatGLM2-6B     & 77.0731 & Qwen2.5-7B-abliterated-v2    & 0.6330     & ChatGLM2-6B     & 56.6711   & SparseLlama-2-7B             & 5.8438 \\
OPT-6.7B        & 118.0234 & \textbf{Quantized Variants} &             & OPT-6.7B        & 108.3884  & \textbf{Quantized Variants}  &  \\
Pythia-6.9B     & 66.2172  & Qwen2.5-7B-4bit             &  0.0041     & Pythia-6.9B     & 102.5610  & Llama2-7B-4bit               & 0.0156 \\
MPT-7B          & 79.5184  & Qwen2.5-7B-8bit             &  0.0041     & MPT-7B          & 44.6246   & Llama2-7B-8bit               & 0.0156 \\
\bottomrule
\end{tabular*}
\label{tab:sim}
\end{table*}

It is important to note that, unlike Llama2-7B, which uses multi-head attention (MHA), Qwen2.5-7B employs Grouped-Query Attention (GQA), which groups multiple query heads to share a single key-value head pair. This architecture introduces a dimensionality mismatch with previous definition in Sec. \ref{sec:attention}, as the number of key and value heads is smaller than that of the query heads. To resolve this compatibility issue, we conceptually broadcast the shared weights in $W_K$ and $W_V$ to align with the number of query heads. This replication functions as a structured weight-sharing constraint, ensuring GQA-based models like Qwen2.5-7B to be seamlessly integrated into our fingerprinting framework.

Table \ref{tab:dist_unrelated} reports the pairwise fingerprint distances between the 10 unrelated models listed in Table \ref{tab:sim}. It can be observed that the distances between the fingerprints of these unrelated models are all greater than 40, exceeding the similarity threshold for unrelated models. This confirms that our method can accurately distinguish unrelated models and thus avoid false positives.

\begin{table*}[h!]
\centering
\renewcommand{\arraystretch}{1.25}
\caption{Fingerprint $L_1$ Distance between Unrelated Models}
\begin{tabular*}{\linewidth}{@{\extracolsep{\fill}}l c c c c c c c c c c}
    \toprule
    & Mis & Qwe & Bai & Int & Gpt & Cer & Cha & OPT & Pyt & MPT \\
    \midrule
Mistral-7B-V0.3   & \      & 40.6418 & 50.7201 & 61.4726 & 64.8229 & 76.6427 & 47.9270 & 119.4732 & 101.0111 & 62.9584 \\
Qwen1.5-7B        & 40.6418 & \      & 48.9473 & 53.5701 & 56.1560 & 75.6061 & 66.0833 & 112.9491 & 84.2514 & 63.4418 \\
Baichuan2-7B      & 50.7201 & 48.9473 & \      & 85.4194 & 53.6894 & 66.7556 & 66.2255 & 97.0414 & 116.8085 & 52.9459 \\
InternLM2.5-7B    & 61.4726 & 53.5701 & 85.4194 & \      & 99.1982 & 119.8395 & 94.6055 & 158.5690 & 46.9688 & 85.8946 \\
GPT2-Large        & 64.8229 & 56.1560 & 53.6894 & 99.1982 & \      & 43.3696 & 72.7921 & 73.1991 & 118.0726 & 52.0082 \\
Cerebras-GPT-1.3B & 76.6427 & 75.6061 & 66.7556 & 119.8395 & 43.3696 & \      & 73.0203 & 60.1043 & 141.7880 & 61.5820 \\
ChatGLM2-6B       & 47.9270 & 66.0833 & 66.2255 & 94.6055 & 72.7921 & 73.0203 & \      & 92.8907 & 132.9924 & 69.6064 \\
OPT-6.7B          & 119.4732 & 112.9491 & 97.0414 & 158.5690 & 73.1991 & 60.1043 & 92.8907 & \      & 167.1513 & 87.0180 \\
Pythia-6.9B       & 101.0111 & 84.2514 & 116.8085 & 46.9688 & 118.0726 & 141.7880 & 132.9924 & 167.1513 & \      & 122.9616 \\
MPT-7B            & 62.9584 & 63.4418 & 52.9459 & 85.8946 & 52.0082 & 61.5820 & 69.6064 & 87.0180 & 122.9616 & \      \\
    \bottomrule
\label{tab:dist_unrelated}
\end{tabular*}
\end{table*}

To validate the effectiveness of our method on distilled models, we further evaluated three DeepSeek \cite{deepseekai2025deepseekr1incentivizingreasoningcapability} distillation models of varying sizes and architectures. As shown in Table \ref{tab:ds}, their distances from respective base models remain less than 5, confirming their relatedness.

% \end{table}

\begin{table}[t]
\centering
\caption{Fingerprint $L_1$ Distance of DeepSeek models}
\begin{tabular*}{\linewidth}{@{\extracolsep{\fill}}llc}
\toprule
Base Model & Distilled Model & Distance \\ \midrule
Qwen2.5-Math-1.5B & DeepSeek-R1-Distill-Qwen-1.5B & 4.7757 \\
Qwen2.5-Math-7B & DeepSeek-R1-Distill-Qwen-7B & 3.3650 \\
Llama-3.1-8B & DeepSeek-R1-Distill-Llama-8B & 4.7858 \\ \bottomrule
\label{tab:ds}
\end{tabular*}
\end{table}

\subsection{Robustness Verification}
\label{sec: robust_veri}

\rev{
\begin{figure*}[t!]
    \centering
    \captionsetup[subfloat]{font={footnotesize}}
    \subfloat[Untargeted Evasion ($L_\text{attack}^U$)\label{fig:ft_untargeted_pure}]{
        \includegraphics[width=0.4\textwidth]{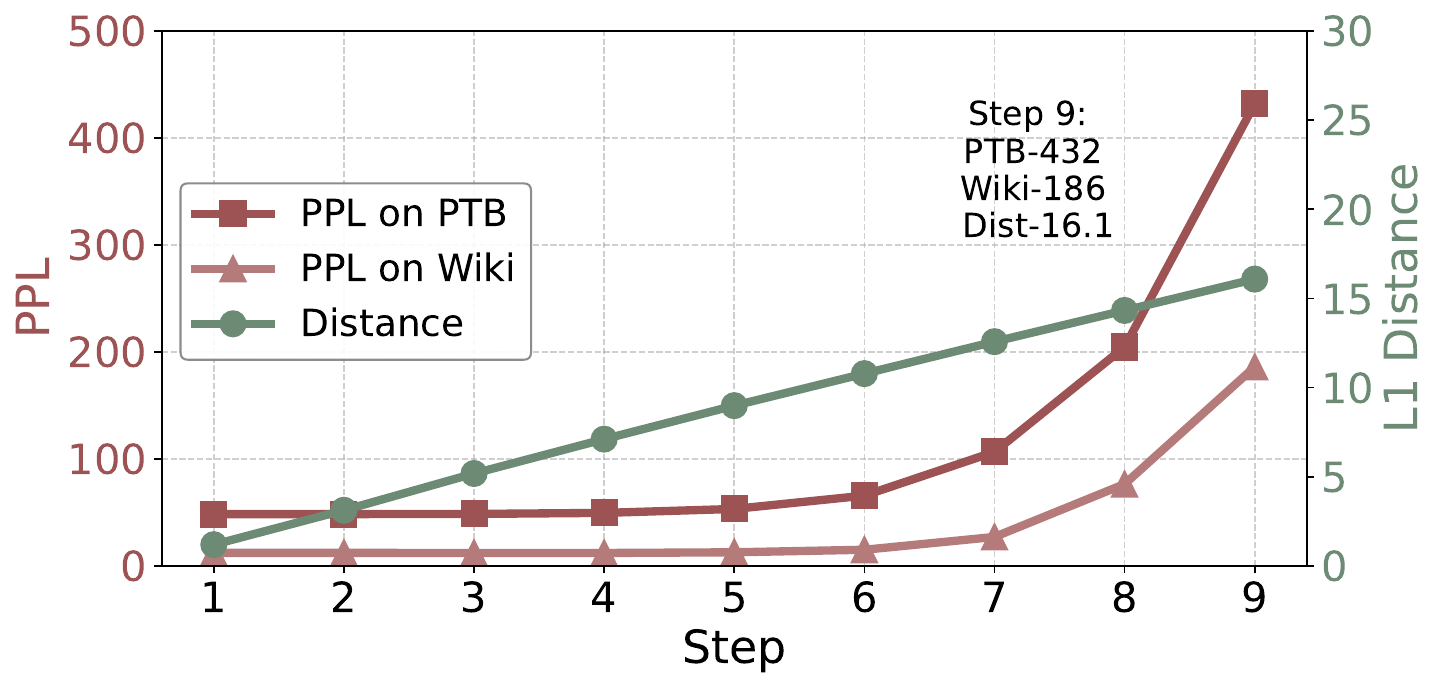}
        \label{fig:ft_attack_u_attack}
    }
    \subfloat[Untargeted Evasion ($L_\text{total}^U$)\label{fig:ft_untargeted_mix}]{
        \includegraphics[width=0.4\textwidth]{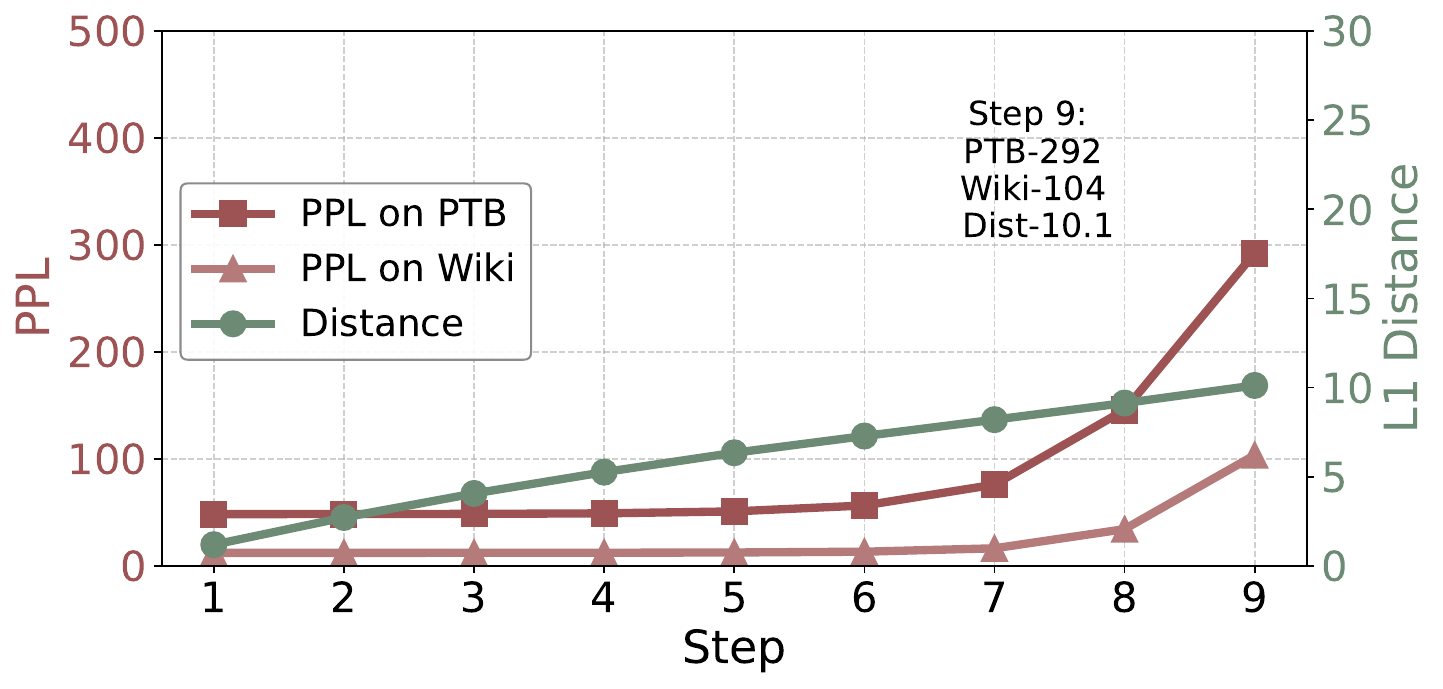}
        \label{fig:ft_attack_u_total}
    }
    \\
    \subfloat[Targeted Evasion ($L_\text{attack}^T$)\label{fig:ft_targeted_pure}]{
        \includegraphics[width=0.4\textwidth]{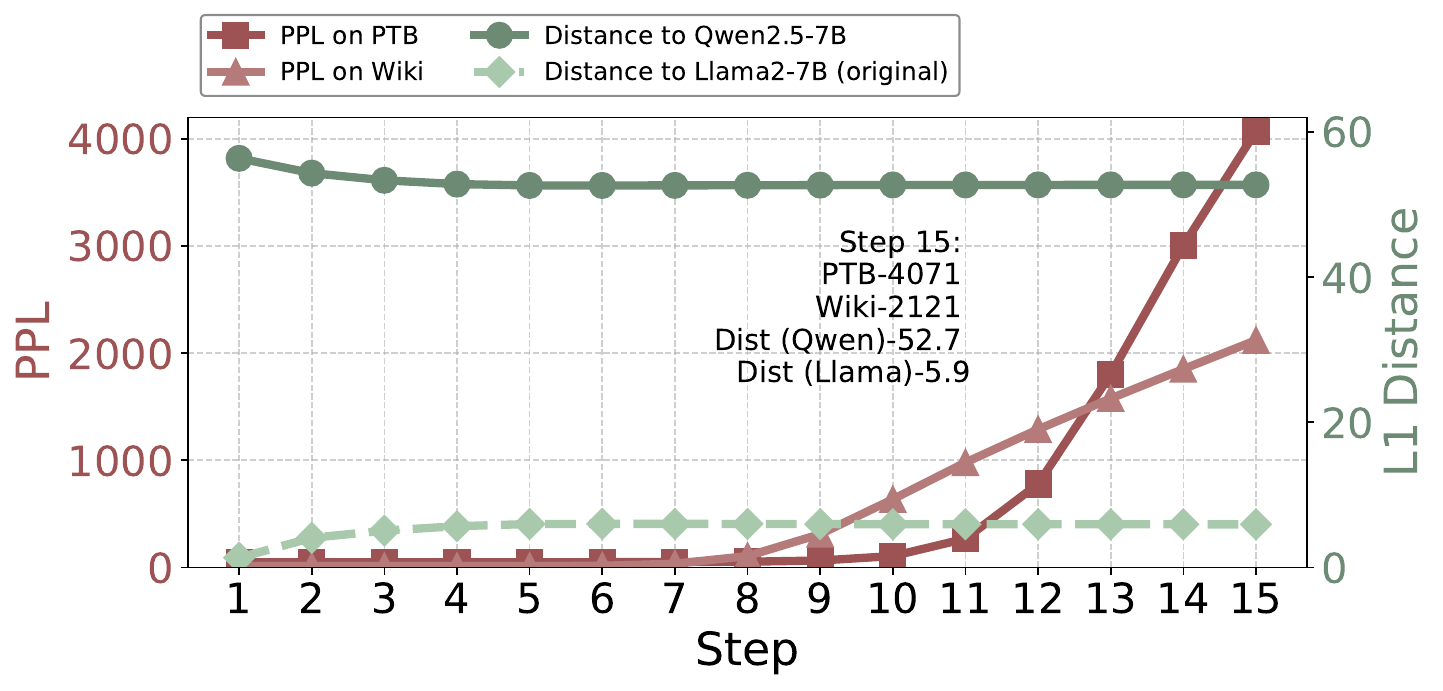}
        \label{fig:ft_attack_t_attack}
    }
    \subfloat[Targeted Evasion ($L_\text{total}^T$)\label{fig:ft_targeted_mix}]{
        \includegraphics[width=0.4\textwidth]{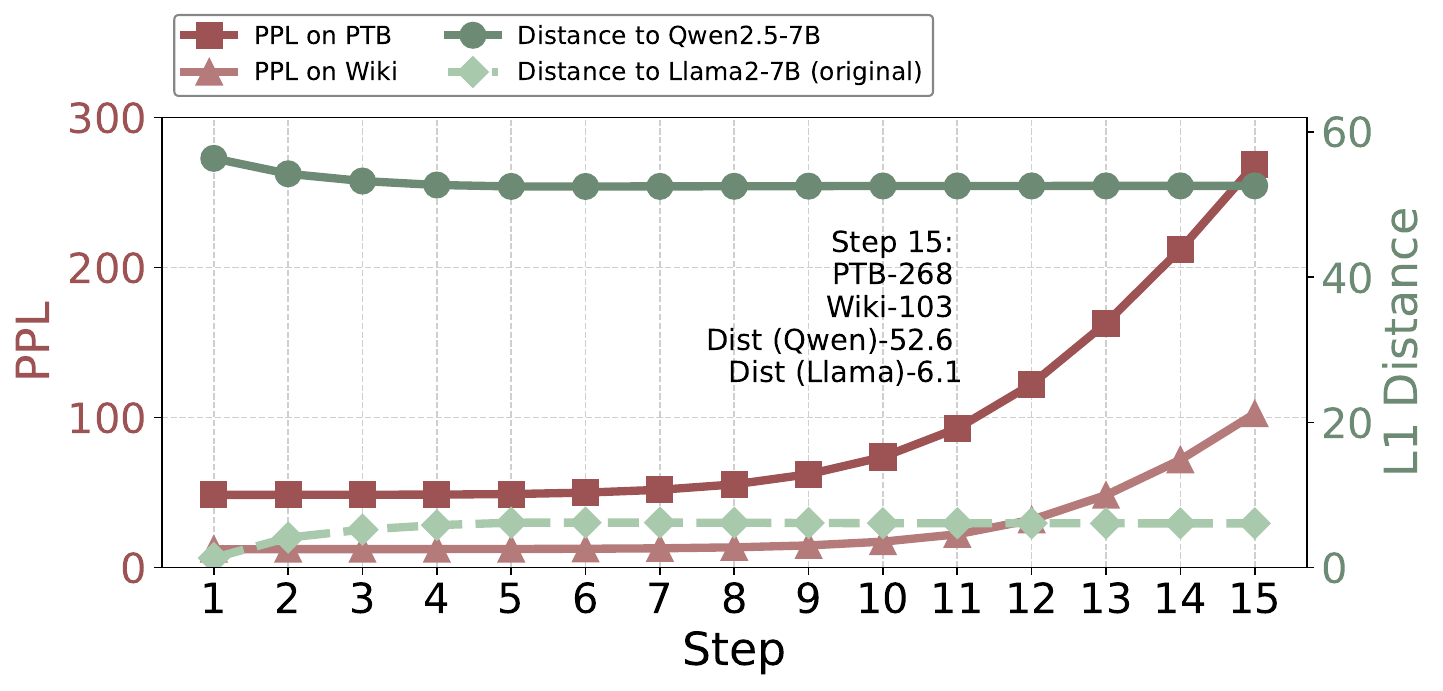}
        \label{fig:ft_attack_t_total}
    }
    \rev{
    \caption{Fingerprint distance and PPL change of Llama2-7B under fine-tuning attacks. (a) and (b) show the results under untargeted evasion attacks. (c) and (d) show the results under targeted evasion attacks. Since SOTA LSTM achieves PPL less than 60 on PTB dataset \cite{merity2018regularizing}, we consider a PPL of 60 or higher as ``unacceptable" for larger models like Transformers.}
    }
    \label{fig:ft_attacks}
\end{figure*}
}

We assess the robustness of SELF by evaluating its resilience against carefully crafted evasion attacks. \rev{Specifically, we investigate scenarios where an adversary attempts to bypass provenance verification by applying model modifications, such as fine-tuning, pruning, and weight transformation.} A robust fingerprint should maintain a high similarity score while preserving model performance under such attacks. \rev{In all experiments, we assume a worst-case white-box adversary who possesses full knowledge of our fingerprint extraction scheme, including the selection of \(W_Q\) and \(W_K\) and the use of the first \(N_\mathcal{F}\) layers.}

\subsubsection{Robustness against Fine-tuning Attack}
\paragraph{\rev{Untargeted attack}} An attacker may fine-tune the \rev{derived} model to evade \rev{provenance} detection by intentionally adjusting model parameters so that the extracted fingerprint $\mathcal{F}_S$ deviates from the original $\mathcal{F}_T$. Formally, this is achieved by optimizing the attack loss 
\rev{
\begin{equation}
     L_\text{attack}^U = \frac{1}{\|\mathcal{F}_{T} - \mathcal{F}_{S}\|_1 + \epsilon} , 
\end{equation}
}
where $\mathcal{F}_S$ is the fingerprint of the suspect model and $\epsilon = 10^{-9}$ serves as a small constant for numerical stability. Moreover, an attacker may also try to use actual dataset to maintain the model's performance, optimizing a combined objective $L_\text{total}$ incorporating both the attack loss and the dataset loss, 
\rev{
\begin{equation}
    L_\text{total}^U=l_1L_\text{attack}^U + l_2L_\text{data}
\end{equation}
} 
where $l_1$ and $l_2$ are balancing hyperparameters. To keep the two loss functions on the same scale, we adopt \(l_1=0.1, l_2=1\). In the experiment, we use the cross-entropy loss on WikiText2 dataset as \(L_\text{data}\). 
% \rev{We assume that the attacker has full knowledge of our fingerprint extraction scheme and thus fine-tunes \(W_Q\) and \(W_K\) in the first \(N_\mathcal{F}\) layers.}
\rev{
\paragraph{Targeted attack} Furthermore, instead of simply diverging from the original model, an attacker might conduct a fingerprint spoofing by altering the derived model's weights to approximate an unrelated model to forge a false identity. In this scenario, the attack loss minimizes the $L_1$ distance to an unrelated model's fingerprint $\mathcal{F}_{U}$:
\begin{equation}
 L_\text{attack}^T = \|\mathcal{F}_{U} - \mathcal{F}_{S}\|_1
\end{equation}
Similar to the untargeted counterpart, this targeted objective can also be combined with $L_\text{data}$ as $L_\text{total}^T$ to maintain utility, using the identical setup.

We implement the above fine-tuning attacks on the Llama2-7B model and select Qwen2.5-7B as the target unrelated model in the targeted attack.} To account for randomness, we average the results from four random seeds (10, 42, 99, 1024).  We analyze both fingerprint similarity and model performance degradation, measuring the latter using perplexity (PPL) on WikiText2 \cite{merity2016pointersentinelmixturemodels} and PTB \cite{marcus-etal-1993-building} benchmarks. A lower PPL indicates better performance. 

Figs. \ref{fig:ft_attack_u_attack} and \ref{fig:ft_attack_u_total} show that even if the model has been extensively fine-tuned to have significantly degraded performance (Step 9), the fingerprint distance to the original still remains low ($<$ \textbf{17}), demonstrating the robustness of our scheme against untargeted fine-tuning attacks. \rev{Similar trends are observed under targeted attacks, where an adversary attempts to simultaneously evade the original lineage and spoof the unrelated target. As shown in Figs. \ref{fig:ft_attack_t_attack} and \ref{fig:ft_attack_t_total}, despite suffering substantial utility loss (e.g., PTB PPL reaches up to 4071 at step 15), the attacked model's fingerprint remains far from Qwen2.5-7B ($\approx$52.7) and tightly bound to the original Llama2-7B ($\approx$6.1). These results show that the evaluated attacks cause substantial utility loss while achieving limited fingerprint divergence from the original model or convergence toward the unrelated target, supporting SELF's robustness against both provenance evasion and fingerprint spoofing.}

% untargeted attacks increase the fingerprint distance from the original model at the cost of substantial performance degradation. Under $L_\text{attack}^U$ and $L_\text{total}^U$, the distances reach 16.1 and 10.1 at step 9, respectively, while PTB PPL rises to 432 and 292, well above the predefined unacceptable threshold. the distances to Qwen2.5-7B decrease only slightly before plateauing at approximately 52.7 and 52.6, while the distances to the original Llama2-7B remain small at approximately 5.9 and 6.1, respectively, even as PTB PPL reaches 4071 and 268 at step 15.

\subsubsection{Robustness against Pruning}
An attacker may attempt to disrupt the fingerprint by increasing pruning levels. Therefore, we examine the impact of varying pruning ratios on fingerprint robustness. Specifically, we prune Llama2-7B with SliceGPT \cite{ashkboos2024slicegpt}, a structured pruning method that removes entire rows or columns from LLM weight matrices while minimizing performance degradation, and analyze fingerprint similarity and performance variations across different pruning levels. 

\begin{figure}[h!]
    \centering
    \includegraphics[width=0.8\columnwidth]{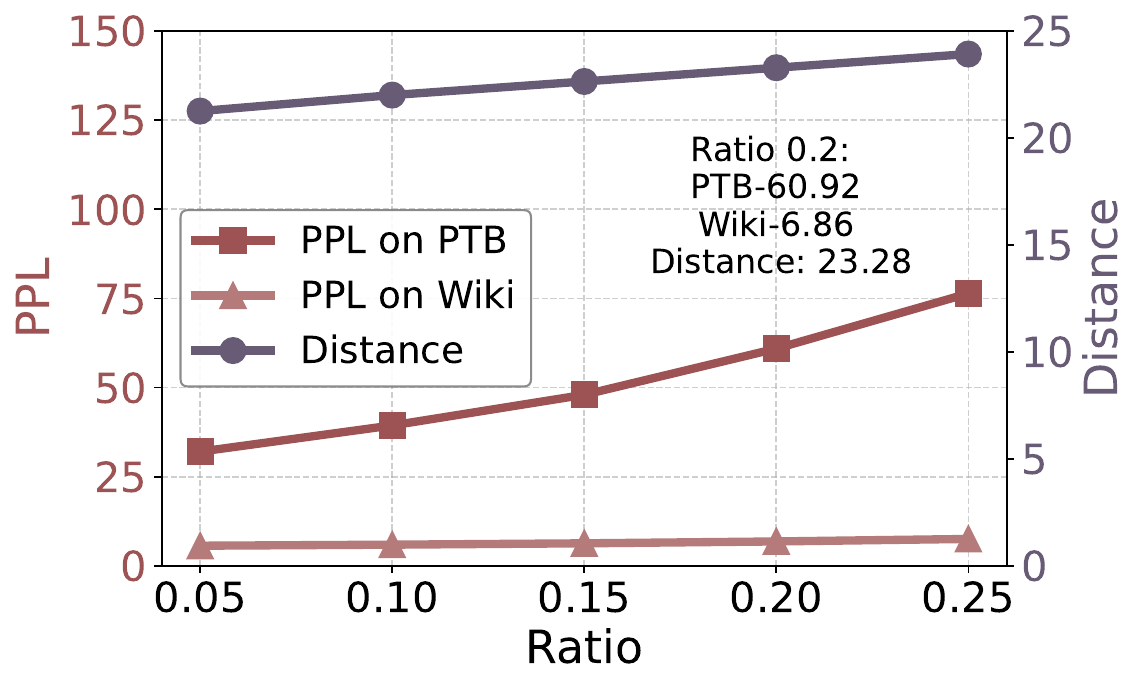}
    \caption{Fingerprint distance and PPL change of Llama2-7B under SliceGPT pruning.}
    \label{fig:pr_atk}
\end{figure}

Fig. \ref{fig:pr_atk} shows that the PPL of Llama2-7B on PTB dataset deteriorates beyond that of an LSTM baseline when the pruning ratio reaches 0.25, but our method maintains a low fingerprint distance ( $<$ 24 ) for infringement detection. We provide more detection results for pruned model instances in Table \ref{tab:compare_s_pr}.

\rev{
\subsubsection{Robustness against Transformation Attack}
We empirically evaluate SELF's robustness to permutation and linear-mapping attacks on OPT-1.3B, complementing the invariance results in Theorems~\ref{sv_invariance} and~\ref{ev_invariance}. Throughout this section, we evaluate the original and transformed models on the same sequence of ten randomly sampled tokens and report the maximum absolute difference in output logits, together with the $L_1$ variations in their query and key weights and extracted fingerprints.

\paragraph{Permutation Attack}
We implement the attack defined in Equation~\ref{eq: per_attack} using a randomly sampled permutation matrix $P\in\{0,1\}^{d_{\mathrm{model}}\times d_{\mathrm{model}}}$. Throughout this description, hidden representations are row vectors, and linear layers follow the convention $y=xW+b$. The same permutation is applied throughout the model, reparameterizing each hidden representation as $x'=xP^\top$.

The token embeddings, positional embeddings, and output projection to vocabulary logits are updated as
\[
E_{\mathrm{tok}}'=E_{\mathrm{tok}}P^\top,\qquad
E_{\mathrm{pos}}'=E_{\mathrm{pos}}P^\top,\qquad
W_{\mathrm{head}}'=PW_{\mathrm{head}},
\]
where $W_{\mathrm{head}}\in\mathbb{R}^{d_{\mathrm{model}}\times |\mathcal{V}|}$ and $|\mathcal{V}|$ denotes the vocabulary size. The scale and bias parameters of every LayerNorm module, represented as row vectors, are reordered as
\[
\gamma'=\gamma P^\top,\qquad
\beta'=\beta P^\top.
\]
For every Transformer layer, the attention and feed-forward weights are permuted as
\[
\begin{aligned}
W_Q' &= PW_Q, &
W_K' &= PW_K, &
W_V' &= PW_V,\\
W_O' &= W_OP^\top, &
W_{\mathrm{fc1}}' &= PW_{\mathrm{fc1}}, &
W_{\mathrm{fc2}}' &= W_{\mathrm{fc2}}P^\top.
\end{aligned}
\]
The query, key, value, and first feed-forward projection biases remain unchanged, whereas the attention output and second feed-forward projection biases are permuted as
\[
b_O'=b_OP^\top,\qquad
b_{\mathrm{fc2}}'=b_{\mathrm{fc2}}P^\top.
\]
These coordinated transformations preserve the model's output logits in exact arithmetic while permuting its hidden representations and corresponding parameters. Table~\ref{tab:permutation_attack} reports the results. The permutation attack substantially alters the query and key weights while producing a logit difference of only $8.21 \times 10^{-5}$. Meanwhile, the fingerprint $L_1$ distance remains as small as $0.0049$, allowing SELF to successfully verify model provenance despite the global reordering of model parameters.

\paragraph{Linear-Mapping Attack}
We implement the attack defined in Equation~\ref{eq: lin_map_attack} using a block-diagonal transformation matrix:
\[
\begin{aligned}
C_1&=\operatorname{diag}\!\left(C_1^{(1)},\ldots,C_1^{(H)}\right)\in\mathbb{R}^{d\times d}, \\
C_1^{(h)}&=G_h+\alpha I_{d_h},\quad h\in[1, \cdots, H]
\end{aligned}
\]
where \(H\) is the number of attention heads,
\(d_h=d_{\mathrm{model}}/H\) is the dimension of each head,
\(G_h\in\mathbb{R}^{d_h\times d_h}\) is sampled element-wise from a
standard Gaussian distribution, and \(I_{d_h}\) denotes the
\(d_h\times d_h\) identity matrix. We set \(\alpha=3.0\) in the
experiments. This diagonal shift moves the singular values of each
random block away from zero in practice, improving the numerical
stability of computing \(C_1^{-1}\) and \(C_1^{-T}\).

For each targeted layer, we transform the query and key projections as
\[
W_Q' = W_Q C_1,
\qquad
W_K' = W_K C_1^{-T},
\]
These updates preserve the attention scores while significantly changing the model weights. We attack the first 8 layers of OPT-1.3B, considering both separate single-layer attacks and a joint attack on all eight layers. Table~\ref{tab:linear_mapping_attack} shows that both the single-layer and joint attacks substantially alter the query and key weights while causing only negligible changes in the output logits. The SELF fingerprint remains stable, with the fingerprint $L_1$ distance not exceeding $0.0259$ across all tested cases, enabling successful provenance verification under the tested linear-mapping attacks.

\rev{
\begin{table}[h]
    \footnotesize
    \centering
    % \color{blue}
    \caption{Experimental Results of the Permutation Attack}
    \label{tab:permutation_attack}
    \vspace{0.1cm}
    \begin{tabular}{cccc}
        \toprule
        \textbf{\makecell{Logit \\ Difference}}
        & \textbf{\makecell{$W_Q$ \\ $L_1$ Distance}}
        & \textbf{\makecell{$W_K$ \\ $L_1$ Distance}}
        & \textbf{\makecell{Fingerprint \\ $L_1$ Distance}} \\
        \midrule
        $8.21 \times 10^{-5}$
        & 2,682,229
        & 2,402,315
        & 0.0049 \\
        \bottomrule
    \end{tabular}
\end{table}
}

\begin{table}[htbp]
    \footnotesize
    \centering
    % \color{blue}
    \caption{Experimental Results of Linear-Mapping Attack on OPT-1.3B}
    \label{tab:linear_mapping_attack}
    \vspace{0.1cm}
    \begin{tabular}{ccccc}
        \toprule
        \textbf{Layer}
        & \textbf{\makecell{Logit \\ Difference}}
        & \textbf{\makecell{$W_Q$ \\ $L_1$ Distance}}
        & \textbf{\makecell{$W_K$ \\ $L_1$ Distance}}
        & \textbf{\makecell{Fingerprint \\ $L_1$ Distance}} \\
        \midrule
        0
        & $9.16 \times 10^{-5}$ & 319,859 & 144,190 & 0.0065 \\
        1
        & $6.77 \times 10^{-5}$ & 525,442 & 208,674 & 0.0032 \\
        2
        & $1.88 \times 10^{-4}$ & 568,894 & 328,554 & 0.0107 \\
        3
        & $1.21 \times 10^{-4}$ & 582,215 & 326,985 & 0.0107 \\
        4
        & $2.68 \times 10^{-4}$ & 611,205 & 329,017 & 0.0071 \\
        5
        & $6.62 \times 10^{-4}$ & 641,005 & 367,997 & 0.0194 \\
        6
        & $1.18 \times 10^{-4}$ & 794,196 & 278,325 & 0.0048 \\
        7
        & $2.21 \times 10^{-4}$ & 780,371 & 321,401 & 0.0031 \\
        \midrule
        0-7
        & $2.55 \times 10^{-4}$ & 4,824,986 & 2,099,348 & 0.0259 \\
        \bottomrule
    \end{tabular}
\end{table}
}
%It can be observed that while the transformation attack introduces significant perturbations to the model weights with negligible impact on performance, our SELF fingerprint remains stable with a remarkably small distance to guarantee successful provenance verification.

\subsection{Qualitative Comparison}

This section benchmarks SELF against published structural fingerprinting methods for LLMs, including \rev{PCS and ICS proposed in HuRef \cite{zeng2024huref}}, and REEF proposed by \cite{zhang2025reef}. A detailed introduction of these related works is presented in Section \ref{app:related_work}.

Table \ref{tab:comparison_mechanism} compares the mechanism and fingerprint size against various modifications across these LLM fingerprinting methods. SELF employs SVD- and EVD- based invariant matrix decomposition, requiring only partial parameters for fingerprint extraction and achieving a compact fingerprint size of approximately \(10^3\) elements (for any model size). Compared to prior works that extract fingerprints ranging from \(10^4\) to \(10^9\) elements (for a 7B model), SELF significantly reduces storage overhead while preserving discriminative capability. 

\begin{table*}[h!]
\footnotesize
\centering
\caption{Comparison of Mechanism and Fingerprint Size}
\label{tab:comparison_mechanism}
\begin{tabular*}{\linewidth}{@{\extracolsep{\fill}}lccc}
\toprule
     & Method & Requirement & Fingerprint size \\ \midrule
\rev{HuRef-PCS}  & Parameter's vector direction   & All parameters               & $\text{parameter\_num} ~ (\sim 10^9)$ \\
\rev{HuRef-ICS}  & Invariant terms' similarity    & Partial parameters and input & $3 \times \text{selected\_layer\_num} \times (\text{sample\_num})^2 ~ (\sim 10^9)$ \\
REEF & Representation's CKA         & Partial parameters and input & $\text{sample\_num} \times d_\text{model} ~ (\sim 10^4 - 10^6)$ \\ \midrule
SELF & Invariant matrix decomposition & Partial parameters           & $2N_{\mathcal{F}} \times h ~ (\sim 10^3)$ \\ \bottomrule
\end{tabular*}
\begin{tablenotes}
    \item CKA: Centered Kernel Alignment
\end{tablenotes}
\end{table*}

\begin{table*}[h!]
\footnotesize
\centering
\caption{Comparison of Robustness Against Various Attacks}
\label{tab:comparison_robustness}
\begin{tabular*}{\linewidth}{@{\extracolsep{\fill}}lccccc}
\toprule
Methods & Model Fine-tuning & Model Pruning & Permutation Attack & Linear-mapping Attack & False Claim Attack \\
\midrule
\rev{HuRef-PCS}  & $\checkmark$ &              &              &              & $\checkmark$ \\
\rev{HuRef-ICS}  & $\checkmark$ &              & $\checkmark$ & $\checkmark$ &              \\
REEF & $\checkmark$ & $\checkmark$ & $\checkmark$ & $\checkmark$ &              \\ \midrule
SELF & $\checkmark$ & $\checkmark$ & $\checkmark$ & $\checkmark$ & $\checkmark$ \\
\bottomrule
\end{tabular*}
\begin{tablenotes}
\item $\checkmark$ indicates resilience against this attack.
\end{tablenotes}
\end{table*}

Table \ref{tab:comparison_robustness} compares robustness against various attacks. By leveraging intrinsic model weights and eliminating input dependency, SELF is inherently robust against false claim attacks. Its SVD- and EVD- constructed fingerprints also provide strong attack resilience against linear mapping and permutation attacks due to their fundamental mathematical invariance properties. Besides, the proposed method is also evaluated to be robust against fine-tuning and pruning attacks, highlighting its effectiveness in practical deployment scenarios. 

\begin{table*}[ht]
\footnotesize
\centering
\caption{Comparison of fingerprint similarity on Llama2-7B fine-tuned models}
\label{tab:compare_ft}
\begin{tabular}{lcccccc}
\toprule
 & \multicolumn{6}{c}{\textbf{Fine-tuning}} \\
\cmidrule(lr){2-7}
 & \shortstack[c]{Llama-2-finance-7B\\(5M Tokens)} 
 & \shortstack[c]{Vicuna-1.5-7B\\(370M Tokens)} 
 & \shortstack[c]{Wizardmath-7B\\(1.8B Tokens)} 
 & \shortstack[c]{Chinesellama-2-7B\\(13B Tokens)} 
 & \shortstack[c]{Codellama-7B\\(500B Tokens)} 
 & \shortstack[c]{Llemma-7B\\(700B Tokens)} \\
\midrule
\rev{HuRef-PCS}  & 0.9979 & 0.9985 & \cellcolor[gray]{0.8}0.0250 & \cellcolor[gray]{0.8}0.0127 & \cellcolor[gray]{0.8}0.0105 & \cellcolor[gray]{0.8}0.0098 \\
\rev{HuRef-ICS}  & 0.9952 & 0.9949 & 0.9994 & \cellcolor[gray]{0.8}0.4996 & \cellcolor[gray]{0.8}0.2550 & \cellcolor[gray]{0.8}0.2257 \\
REEF & 0.9950 & 0.9985 & 0.9979 & 0.9974 & 0.9947 & 0.9962 \\ \midrule
% SELF (Distance) & 0.0420 & 0.0393 & 0.0226 & 0.0576 & 0.6358 & 0.5739 \\
SELF (Distance) & 0.8504 & 0.8598 & 0.4559 & 1.2147 & 12.6612 & 12.7050 \\
\bottomrule
\end{tabular}
\end{table*}

\begin{table*}[h!]
\footnotesize
\centering
\caption{Comparison of fingerprint similarity on Llama2-7B structured pruned models}
\label{tab:compare_s_pr}
\begin{tabular}{lcccccc}
\toprule
 & \multicolumn{6}{c}{\textbf{Structured Pruning}} \\
\cmidrule(lr){2-7}
 & \shortstack[c]{Sheared-Llama-\\1.3B-pruned} 
 & \shortstack[c]{Sheared-Llama-\\1.3B} 
 & \shortstack[c]{Sheared-Llama-\\1.3B-sharegpt} 
 & \shortstack[c]{Sheared-Llama-\\2.7B-pruned} 
 & \shortstack[c]{Sheared-Llama-\\2.7B} 
 & \shortstack[c]{Sheared-Llama-\\2.7B-sharegpt} \\
\midrule
\rev{HuRef-PCS}  & \cellcolor[gray]{0.8}0.0000 & \cellcolor[gray]{0.8}0.0000 & \cellcolor[gray]{0.8}0.0000 & \cellcolor[gray]{0.8}0.0000 & \cellcolor[gray]{0.8}0.0000 & \cellcolor[gray]{0.8}0.0000 \\
\rev{HuRef-ICS}  & \cellcolor[gray]{0.8}0.4927 & \cellcolor[gray]{0.8}0.3512 & \cellcolor[gray]{0.8}0.3510 & 0.6055 & \cellcolor[gray]{0.8}0.4580 & \cellcolor[gray]{0.8}0.4548 \\
REEF & 0.9368 & 0.9676 & 0.9710 & 0.9278 & 0.9701 & 0.9991 \\ \midrule
SELF (Distance) & 23.2506 & 16.8061 & 16.7971 & 11.0471 & 9.8122 & 11.0611 \\
\bottomrule
\end{tabular}
\begin{tablenotes}
    \item \textit{-pruned}: the version without continued pre-training; 
    \item \textit{-sharegpt}: the version with instruction tuning.
\end{tablenotes}
\end{table*}

\begin{table}[ht]
\footnotesize
\centering
\caption{Comparison of fingerprint similarity score on Llama2-7B unstructured pruned models}
\begin{tabular}{lccc}
\toprule
\multicolumn{4}{c}{\textbf{Unstructured Pruning}} \\
\cmidrule(lr){2-4}
 & \shortstack[c]{Sparse-Llama-\\2-7B} & \shortstack[c]{Wanda-Llama-\\2-7B} & \shortstack[c]{GBLM-Llama-\\2-7B} \\
\midrule
\rev{HuRef-PCS} & 0.9560 & 0.9620 & 0.9616 \\
\rev{HuRef-ICS} & 0.9468 & 0.9468 & 0.9478 \\
REEF & 0.9985 & 0.9986 & 0.9991 \\ \midrule 
SELF (Distance) &  2.7556 & 3.0326 & 3.1229  \\
\bottomrule
\end{tabular}
\label{tab:compare_u_pr}
\end{table}

\subsection{Quantitative Comparison}
\label{sec: quan_compare}

To further demonstrate the numerical superiority of our method, this section presents quantitative comparisons between our method and previous approaches. Following \cite{zhang2025reef}, we present the IP detection results on publicly available models in Table \ref{tab:compare_ft} to Table \ref{tab:compare_merge}. The performance (similarity) for previous methods are sourced directly from \cite{zhang2025reef}, which conducted comprehensive comparison with existing works and reported the corresponding data. For consistency, we reuse these reported values to benchmark our method against the SOTA works. False judgments are highlighted in gray and the decision threshold for SELF is set as $\tau =$\textbf{24} for comparison.

\paragraph{Comparison of fine-tuned variants detectability}
Table \ref{tab:compare_ft} compares SELF with existing methods for detecting fine-tuned related models. The target model is Llama2-7B and its six fine-tuned variants are set as suspect models, fine-tuned on datasets ranging from 5M to 700B tokens. It can be observed that even after the target model is fine-tuned on large-scale data, SELF still identifies related models with a distance less than 13, while HuRef-PCS and HuRef-ICS failed to detect some fine-tuned related models.

\paragraph{Comparison of pruned variants detectability}
Tables \ref{tab:compare_s_pr} and \ref{tab:compare_u_pr} compare SELF with existing methods for detecting pruned related models. Llama-2-7B is used as the target model and its pruned variants as suspect models. Table \ref{tab:compare_s_pr} lists the fingerprint similarity scores of Llama-2-7B's structured pruned models and Table \ref{tab:compare_u_pr} reports those of unstructured pruned models. The results show that pruned related models all exhibit fingerprint distances below the predefined threshold, thereby demonstrating the successful detection of model similarity.

\paragraph{Comparison of merged models detectability}
\label{par:reef_fp}

In the model merging scenario, a suspect model is often derived from multiple victim models' weights. Table \ref{tab:compare_merge} presents the fingerprint similarity scores between weight-merged model (EvoLLM-jp-7B) and its corresponding victim target models. The results demonstrate that SELF reliably detects weight merged models as related.

\begin{table}[ht!]
\footnotesize
\centering
\caption{Comparison of fingerprint similarity between EvoLLM-jp-7B and its parent models}
\label{tab:compare_merge}
\begin{tabular}{lccc}
\toprule
  & \begin{tabular}{@{}c@{}}Shisa-gamma-\\7B-v1\end{tabular} 
  & \begin{tabular}{@{}c@{}}Wizard-math-\\7B-1.1\end{tabular} 
  & \begin{tabular}{@{}c@{}}Abel-\\7B-002\end{tabular} \\
\midrule
\rev{HuRef-PCS} & 0.9992 & 0.9990 & 0.9989 \\
\rev{HuRef-ICS} & 0.9992 & 0.9988 & 0.9988 \\
REEF & 0.9635 & 0.9526 & 0.9374 \\ \midrule
SELF (Distance) & 2.1069  & 2.1487  & 2.1867 \\
\bottomrule
\end{tabular}
\end{table}

\paragraph{Comparison on large-scale benchmark} In addition, we evaluated and compared SELF with HuRef-ICS and REEF (HuRef-PCS is omitted as its susceptibility to transformation attacks limits its practical application) on LeafBench \cite{shao2025soklargelanguagemodel}, \rev{a benchmark covering 65 model instances from seven base-model lineages: Qwen2.5-7B, Qwen2.5-14B, Llama-3.1-8B, Mistral-7B-v0.3, Gemma-2-2B, TinyLlama-1.1B, and Llama-2-7B. These lineages span multiple model families, both MHA and GQA, and parameter scales ranging from 1.1B to 14B.} Models with fingerprint distances below the threshold \(\tau = 24\) are considered related, and those above are unrelated. Table \ref{tab:compare_lb} reports \rev{the Area Under the Curve (AUC), detection Accuracy (ACC), True Positive Rate (TPR) at a 1\% False Positive Rate (FPR), and Mahalanobis Distance (MD)}. The results demonstrate that SELF achieves state-of-the-art performance on LeafBench, outperforming existing HuRef-ICS and REEF approaches. The main failure cases for SELF on LeafBench involve dot-v2.7 in TinyLlama-1.1B-Chat-v1.0 family. Although dot-v2.7 merges weights from TinyLlama, SELF fails to recognize the relationship due to the minimal integration ratio. Notably, other methods also failed to identify this instance.

\rev{
\begin{table}[h!]
\footnotesize
\centering
%\color{blue}
\caption{Comparison of Evaluation on LeafBench}
\begin{tabular}{lccccc}
\toprule
Method  & AUC    & ACC   & TPR@1\%FPR  & MD \\
\midrule
\rev{HuRef-ICS}     & \textbf{0.987}  & 0.983 & 0.888      & 2.547 \\
REEF    & 0.878  & 0.894  & 0.627  & 1.993 \\ \midrule
SELF    & 0.986  & \textbf{0.996}  & \textbf{0.979} & \textbf{2.585 } \\
\bottomrule
\end{tabular}
\label{tab:compare_lb}
\end{table}
}

\rev{
\begin{figure*}[t]
    \centering
    \captionsetup[subfloat]{font={footnotesize}}
    \subfloat[$L_1$ Distance \label{fig:lb_metric_l1}]{
        \includegraphics[width=0.31\textwidth]{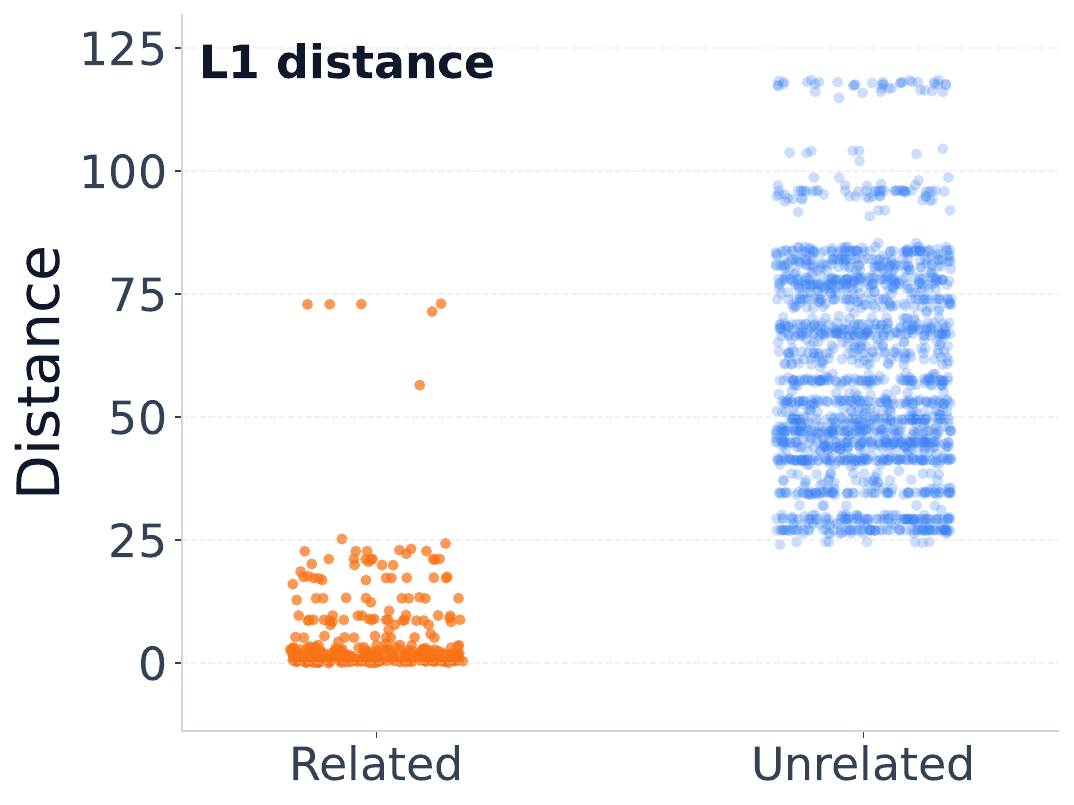}
    }
    \hfill
    \subfloat[$L_2$ Distance \label{fig:lb_metric_l2}]{
        \includegraphics[width=0.31\textwidth]{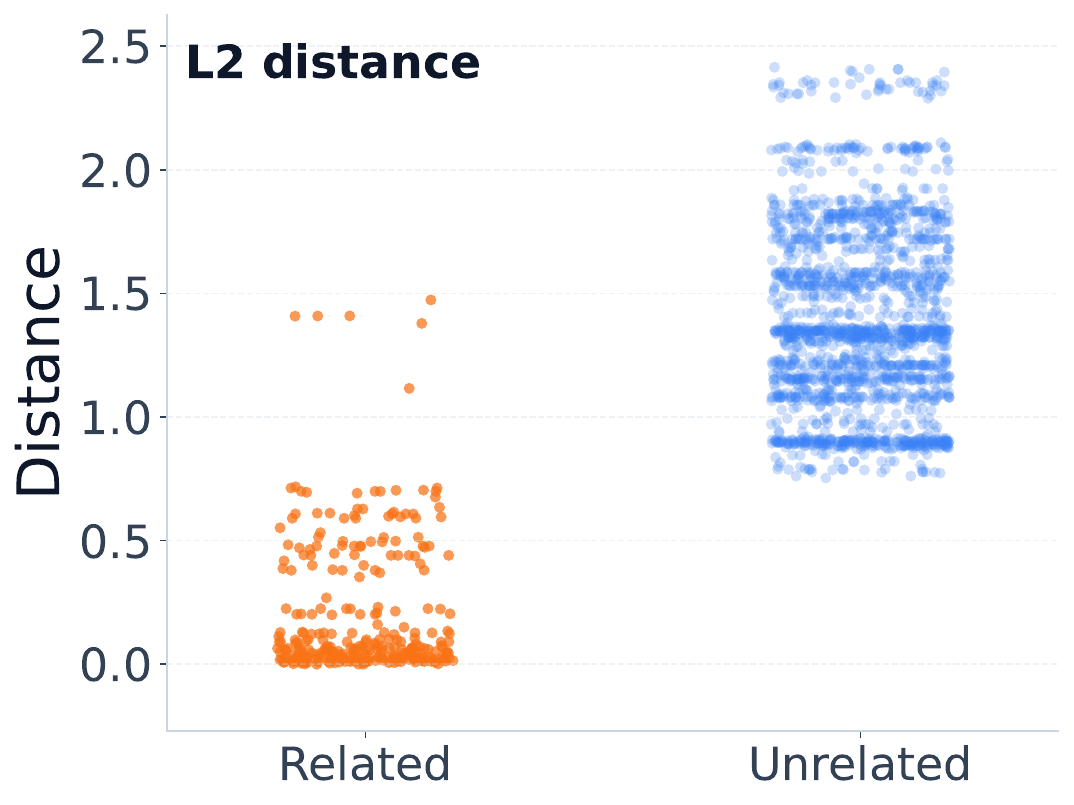}
    }
    \hfill
    \subfloat[Cosine Distance\label{fig:lb_metric_cosine}]{
        \includegraphics[width=0.31\textwidth]{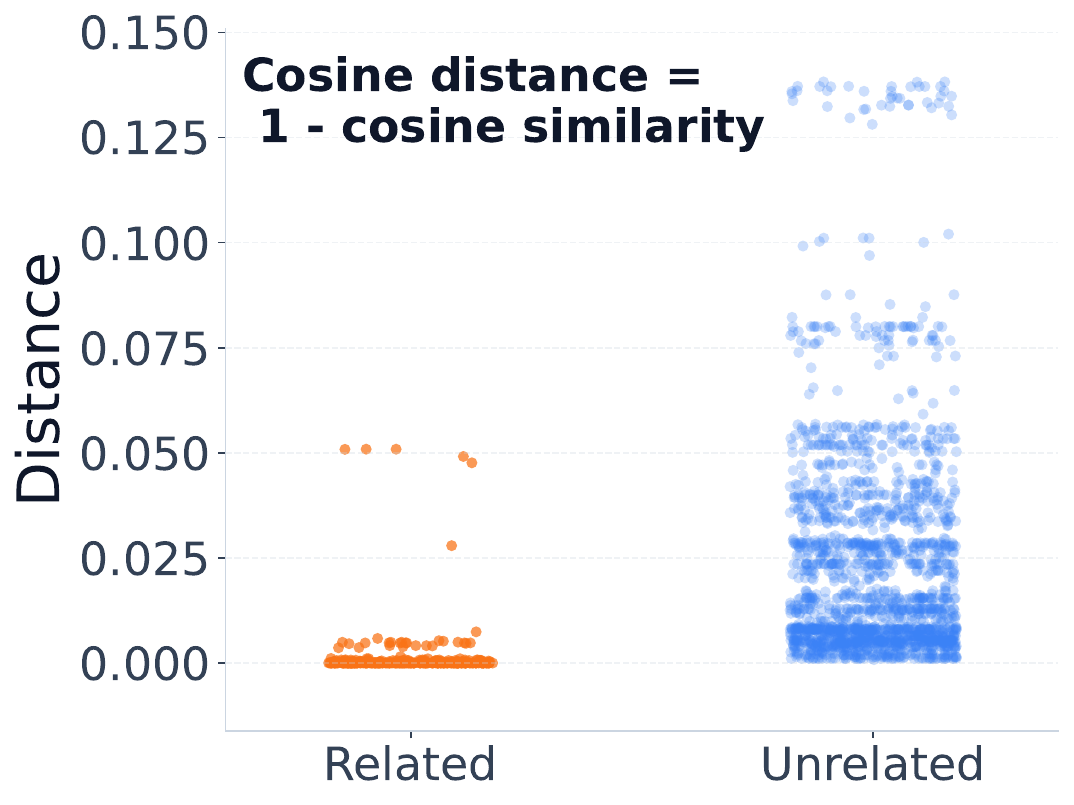}
    }
    \rev{\caption{Distance Distribution with Different Metrics on LeafBench}
    \label{fig:lb_metrics}}
\end{figure*}
}

\subsection{Ablation Studies}
\subsubsection{Distance Metric Ablation and Threshold Sensitivity}
\label{sec: lb_ablation}
\rev{
This section evaluates the selection of distance metric and the sensitivity of decision threshold using the 65-instance LeafBench dataset. 
To empirically validate our metric selection, we compared our default $L_1$ distance against two commonly used alternatives: $L_2$ distance and cosine distance ($1 - \text{cosine similarity}$). Fig.~\ref{fig:lb_metrics} illustrates the resulting distance distributions for related and unrelated models. Although both $L_1$ and $L_2$ distances prove effective, we adopt the $L_1$ distance to align with singular value and eigenvalue theory in Section \ref{sec:sv_ev}.
}

\begin{figure}[t]
    \centering
    \includegraphics[width=0.9\linewidth]{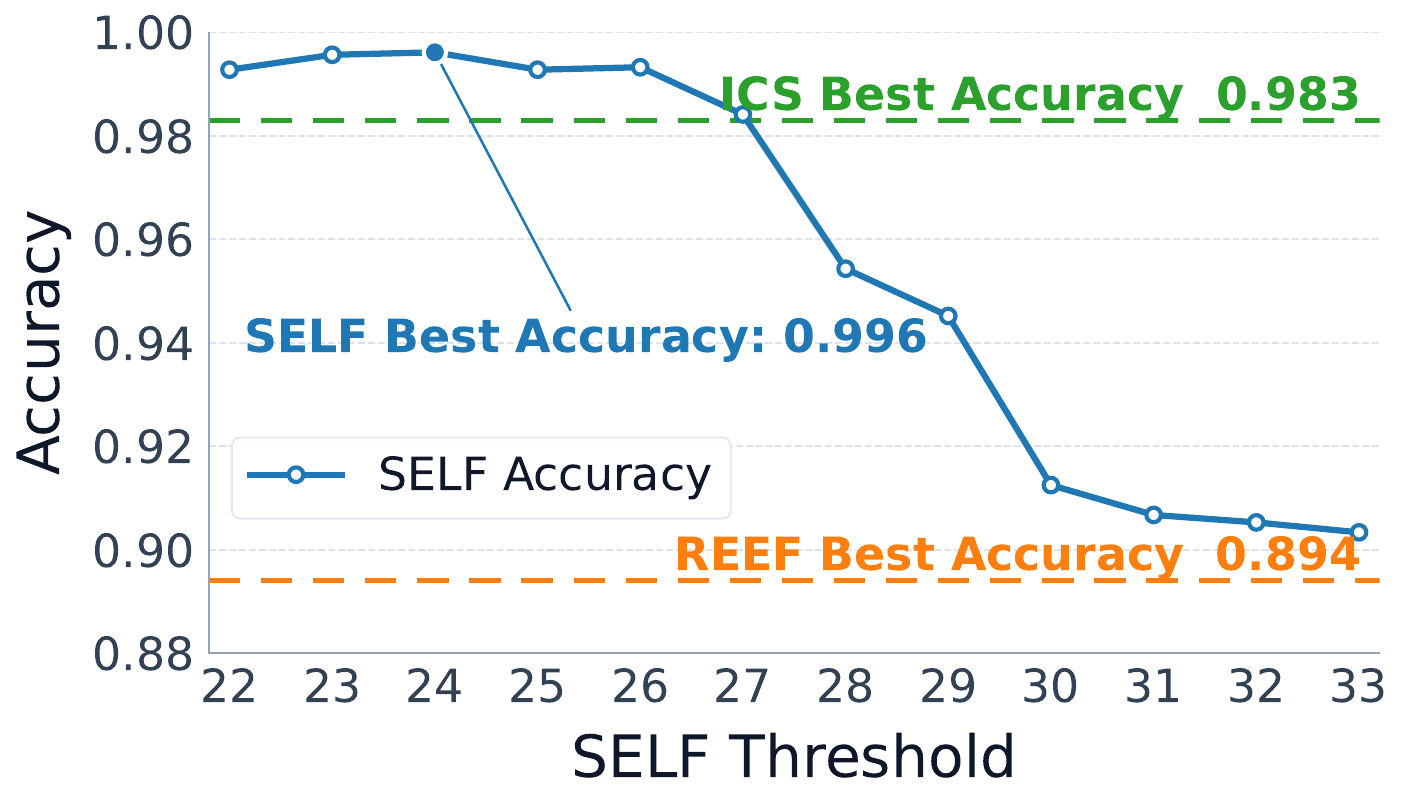}
    \rev{\caption{Accuracy on LeafBench with Different Thresholds}
    \label{fig:abs_tau}}
\end{figure}

\rev{Furthermore, to evaluate the sensitivity of the SELF method to the threshold $\tau$, we varied $\tau$ across the critical decision margin from 22 to 33 based on the results in Table \ref{tab:sim}. Fig. \ref{fig:abs_tau} shows the corresponding accuracy variations on LeafBench. SELF achieves its highest accuracy of 99.6\% at $\tau=24$, motivating our choice of this threshold. For $\tau \in [22,26]$, SELF maintains an accuracy above 99\%, indicating limited sensitivity to small threshold changes around the selected value, while consistently outperforming the best accuracy of both HuRef-ICS (98.3\%) and REEF (89.4\%).}

\subsubsection{Ablation on Fingerprint Extraction Settings}
In this section, we assess how different settings affect fingerprint extraction. We define the \textbf{fingerprint margin} as the difference between: 1) the minimum $L_1$ distance among unrelated models' fingerprints, and 2) the maximum $L_1$ distance among related models' fingerprints. Here, we have considered all models appeared in Tables \ref{tab:sim} to \ref{tab:compare_u_pr}. A large positive margin indicates clearer distinguishability between related and unrelated models, while a negative margin suggests  that some unrelated models exhibit smaller fingerprint distances than the maximum observed among related models. Accordingly, we systematically investigate the impact of the following settings on the fingerprint margin: 
\begin{itemize}
\item Invariant matrix construction choice (singular values or eigenvalues)
\item Transformer block layers selection (e.g. early vs. late layers) 
\item Weight selection (e.g. \(W_Q\) and \(W_K\) vs. \(W_V\) and \(W_O\))
\item Parameter \(h\) (top-\(h\) singular values / eigenvalues are selected to form the fingerprint)
\item Parameter \(N_{\mathcal{F}}\) (how many layers are selected to form the fingerprint)
\end{itemize}

As shown in Table \ref{tab:ablation_layer}, the first eight layers offer larger fingerprint margins, indicating stronger differentiation between related and unrelated models than other layers. 

\begin{table}[h]
\footnotesize
\centering
\caption{Fingerprint Margin with Different Layer Selections}
\label{tab:ablation_layer}
\begin{tabular}{lccc}
\toprule
\textbf{Invariant Matrix} & \textbf{First 8} & \textbf{Middle 8} & \textbf{Last 8} \\
\midrule
Singular Value & 3.7414  & 0.7141  & -5.3285 \\
Eigenvalue     & 7.5367  & -1.6868 & -7.0671 \\
Both           & 11.2780 & -0.9727 & -10.3403\\
\bottomrule
\end{tabular}
\end{table}

\begin{table}[h]
\footnotesize
\centering
\caption{Fingerprint Margin with Different Weight Selections}
\label{tab:ablation_weight_select}
\begin{tabular}{lccc}
\toprule
\textbf{Invariant Matrix} & $W_Q, W_K$ & $W_V, W_O$ & \textbf{Both} \\
\midrule
Singular Value & 3.7414  & -3.5293 & 1.6019 \\
Eigenvalue     & 7.5367  & -6.5116 & 6.1064 \\
Both           & 11.2780 & -8.0503 & 7.7083 \\
\bottomrule
\end{tabular}
\end{table}

We further investigate the performance impact of selecting different weight sources for fingerprint extraction: the pair (\(W_Q\) and \(W_K\)), the pair (\(W_V\) and \(W_O\)) or a combination of both. Similar to the approach taken with \(W_Q\) and \(W_K\), when choosing \(W_V\) and \(W_O\) as the fingerprint source, we formulated the potential attacks as 
\begin{equation}
    \hat{W}_V = P W_V C_2, \quad \hat{W}_O = C_2^{-1} W_O P^T.
    \label{eq:attack_form_wvo}
\end{equation}
The singular value invariant matrix and eigenvalue invariant matrix can be constructed as 
\begin{equation}
\begin{split}
    &Y_\sigma = W_V W_O \in \mathbb{R}^{d_{\text{model}} \times d_{\text{model}}}, \\
    &Y_\lambda = W_O W_V \in \mathbb{R}^{d \times d},
\end{split}
    \label{eq:sigma_matrix_wvo}
\end{equation}
since
\begin{equation}
\begin{split}
        \hat{Y}_\sigma & = (P W_V C_2)(C_2^{-1} W_O P^T)  \\ & = P W_V C_2 C_2^{-1} W_O P^T \\ &= P W_V W_O P^T = P Y_\sigma P^T, \\
        \hat{Y}_\lambda & = (C_2^{-1} W_O P^T)(P W_V C_2) \\ & = C_2^{-1} W_O P^T P W_V C_2 \\ &= C_2^{-1} W_O W_V C_2 = C_2^{-1} Y_\lambda C_2,
\end{split}
\end{equation}
which ensures invariant singular values and eigenvalues under the transformation attacks. Then we calculate the normalized top singular values and eigenvalues from $Y_\sigma$ and $Y_\lambda$, respectively, to form the fingerprint. The results in Table \ref{tab:ablation_weight_select} indicate that selecting \(W_Q\) and \(W_K\) yields the best discriminative ability for SELF.

% This superiority stems from the distinct functional roles of these matrices \cite{elhage2021mathematical}. Specifically, \(W_Q\) and \(W_K\) are primarily responsible for attention routing, whereas \(W_V\) and \(W_O\) function as content-writing mechanisms that determine how an attended token influences the output logits. Because downstream modifications predominantly adapt a model's factual knowledge and stylistic behaviors rather than its core routing logic, \(W_Q\) and \(W_K\) undergo significantly less modification. Consequently, their derived invariants exhibit higher stability for fingerprint extraction.

\begin{table}[h!]
\footnotesize
\centering
\caption{Fingerprint Margin under Varying $h$}
\label{tab:ablation_h}
\begin{tabular}{lcccc}
\toprule
\textbf{Invariant Matrix} & \textbf{32} & \textbf{64} & \textbf{128} & \textbf{256} \\
\midrule
Singular Value & -0.1411 & 0.5282 & 2.5339 & 3.7414  \\
Eigenvalue     & -1.0599 & 0.9153 & 4.2879 & 7.5367  \\
Both           & -1.0805 & 2.3152 & 7.8773 & 11.2780 \\
\bottomrule
\end{tabular}
\end{table}

\begin{table}[h!]
\footnotesize
\centering
\caption{Fingerprint Margin under Varying $N_{\mathcal{F}}$}
\label{tab:ablation_nf}
\begin{tabular}{lcccc}
\toprule
\textbf{Invariant Matrix} & \textbf{2} & \textbf{4} & \textbf{6} & \textbf{8} \\
\midrule
Singular Value & -1.6139 & 1.9465 & 3.4775 & 3.7414  \\
Eigenvalue     & -2.3803 & 1.1200 & 4.9059 & 7.5367  \\
Both           & -3.4706 & 3.0666 & 8.9341 & 11.2780 \\
\bottomrule
\end{tabular}
\end{table}

Tables \ref{tab:ablation_h} and \ref{tab:ablation_nf} confirm that even small \(h\) and \(N_\mathcal{F}\) can provide discriminative capability, while larger \(h\) and \(N_\mathcal{F}\) lead to a greater fingerprint margin. The selection of \(h\) should ensure robust fingerprint extraction while being practical for deployment across diverse model sizes, i.e., smaller than any model's \(d\) and \(d_{\text{model}}\). Thus, we chose \(h=256\) and \(N_\mathcal{F}=8\) in our experiments, which is compatible with most LLMs' weight dimensions.

\rev{\subsection{False-claim Vulnerabilities in Existing Methods}
\label{sec: false_claim}
}

\rev{Input-dependent fingerprinting methods may be vulnerable to false-claim attacks, where an attacker exploits inputs to artificially inflate the similarity between unrelated models without modifying their parameters. We investigate this vulnerability in REEF and HuRef-ICS: the former's heavy reliance on evaluation datasets naturally triggers severe false positives; whereas the latter allows active exploitation via adversarially crafted input tokens.}

\begin{table*}[h!]
\footnotesize
\centering
\caption{Comparison of Llama2-7B unrelated models fingerprint similarity \rev{with different input datasets}}
\label{tab:reef_unrelated}
\begin{tabular}{lcccccccc}
\toprule
Method & Dataset & Mistral-7B & MPT-7B & InternLM2.5 & GPT2-Large & Cerebras-GPT & Pythia-6.9B & OPT-6.7B \\
\midrule
\multirow{5}{*}{REEF} 
  & TruthfulQA   & 0.1975 & 0.2111 & 0.1942 & 0.2320 & 0.2257 & 0.2307 & 0.2437 \\
  & ConfAIde     & 0.2340 & 0.2449 & 0.2308 & 0.2464 & 0.2471 & 0.2723 & 0.2419 \\
  & PKU-SafeRLHF & \cellcolor[gray]{0.8}0.5099 & \cellcolor[gray]{0.8}0.5150 & 0.4307 & 0.4808 & 0.4390 & 0.4844 & 0.4764 \\
  & ToxiGen& \cellcolor[gray]{0.8}0.5894 & \cellcolor[gray]{0.8}0.5528 & \cellcolor[gray]{0.8}0.6067 & \cellcolor[gray]{0.8}0.5366 & \cellcolor[gray]{0.8}0.6358 & \cellcolor[gray]{0.8}0.6151 & 0.4716 \\
  & SST2  & \cellcolor[gray]{0.8}0.6772 & \cellcolor[gray]{0.8}0.6221 & \cellcolor[gray]{0.8}0.6628 & \cellcolor[gray]{0.8}0.6034 & \cellcolor[gray]{0.8}0.5945 & \cellcolor[gray]{0.8}0.6783 & \cellcolor[gray]{0.8}0.5982 \\
\midrule
\multicolumn{2}{c}{SELF (Distance)} & 34.5286 & 44.6246 & 65.0354 & 49.9396 & 67.3757 & 102.5610 & 108.3884 \\
\bottomrule
\end{tabular}
\end{table*}

\rev{
\paragraph{False-claim risk in REEF}
% Although REEF achieves state-of-the-art performance in Tables \ref{tab:compare_ft}--\ref{tab:compare_merge}, its dependence on the evaluation inputs can lead to high similarity between unrelated models, increasing the risk of false positives and input-driven false claims.
To empirically demonstrate REEF's passive vulnerability, we compare Llama2-7B against seven unrelated models (Table \ref{tab:reef_unrelated}) using five diverse datasets: 
TruthfulQA \cite{lin2022truthfulqa},
ConfAIde \cite{mireshghallahcan},
PKU-SafeRLHF \cite{ji2025pkusaferlhfmultilevelsafetyalignment},
ToxiGen \cite{hartvigsen2022toxigen}, and
SST2 \cite{socher2013recursive}.
Following the original protocol in \cite{zhang2025reef}, we evaluate REEF using representation-based fingerprinting, specifically CKA at the 18th layer, across 200 samples per dataset.

As shown in Table \ref{tab:reef_unrelated}, REEF's similarity scores for unrelated models fluctuate drastically across inputs, peaking at \textbf{0.6783} on SST2. This severe input-driven volatility practically obliterates the decision margin, enabling an attacker to effortlessly cherry-pick evaluation datasets to fabricate a false ownership claim. 
In contrast, our SELF method maintains consistently high distances ($>$34) for unrelated models, demonstrating superior robustness against false claim attacks.
}

\begin{figure*}[t]
    \centering
    \includegraphics[width=0.8\linewidth]{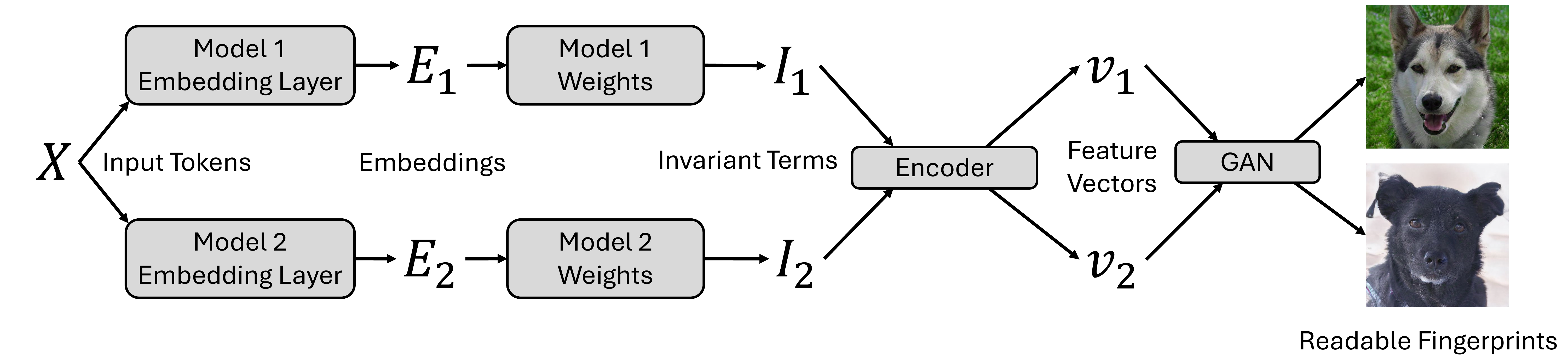}
    \caption{HuRef calculation process in our false claim attack.}
    \label{fig:huref_compair}
\end{figure*}

\begin{figure*}[t]
    \centering
    \captionsetup[subfloat]{font=footnotesize}
    \subfloat[Feature vectors' similarity change]{
        \includegraphics[width=0.32\linewidth]{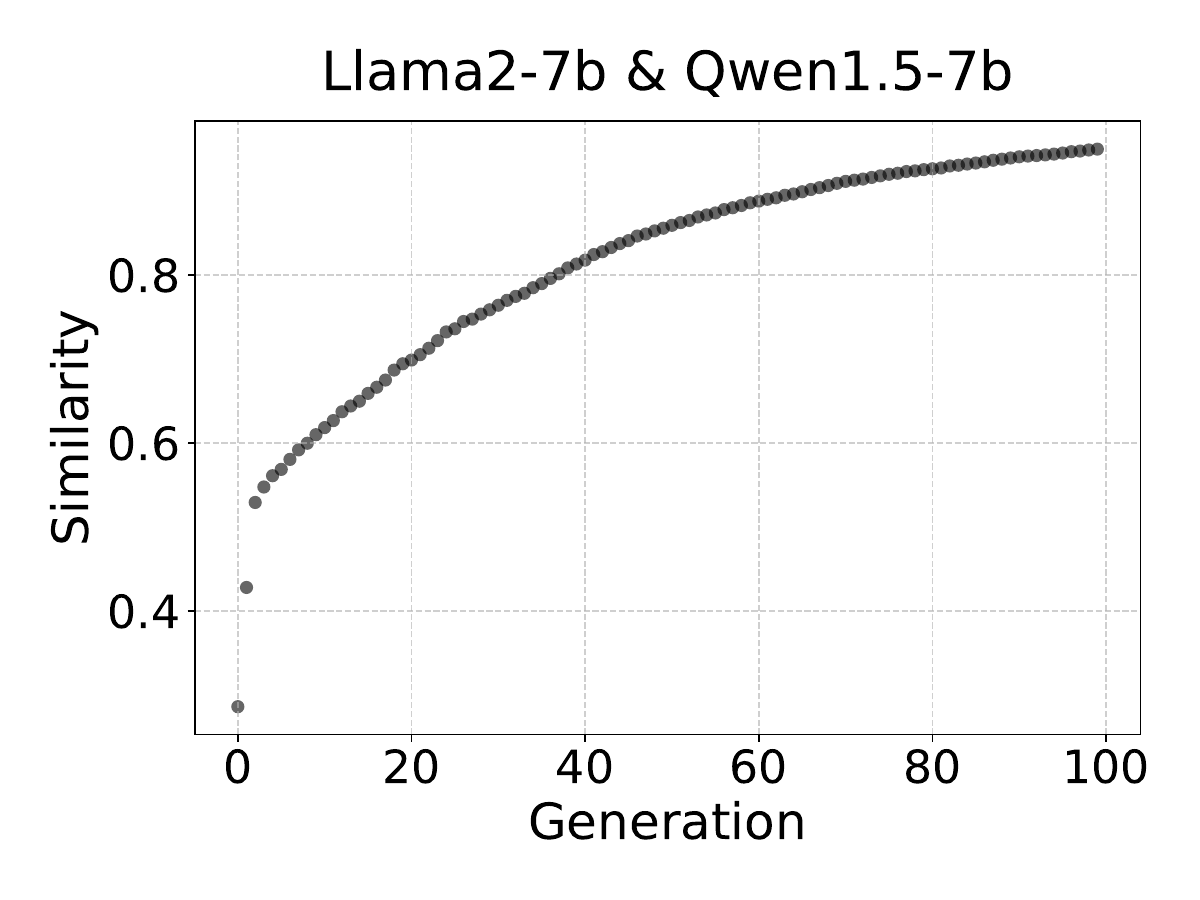}
    }
    \subfloat[Human-readable fingerprint change]{
        \includegraphics[width=0.32\linewidth]{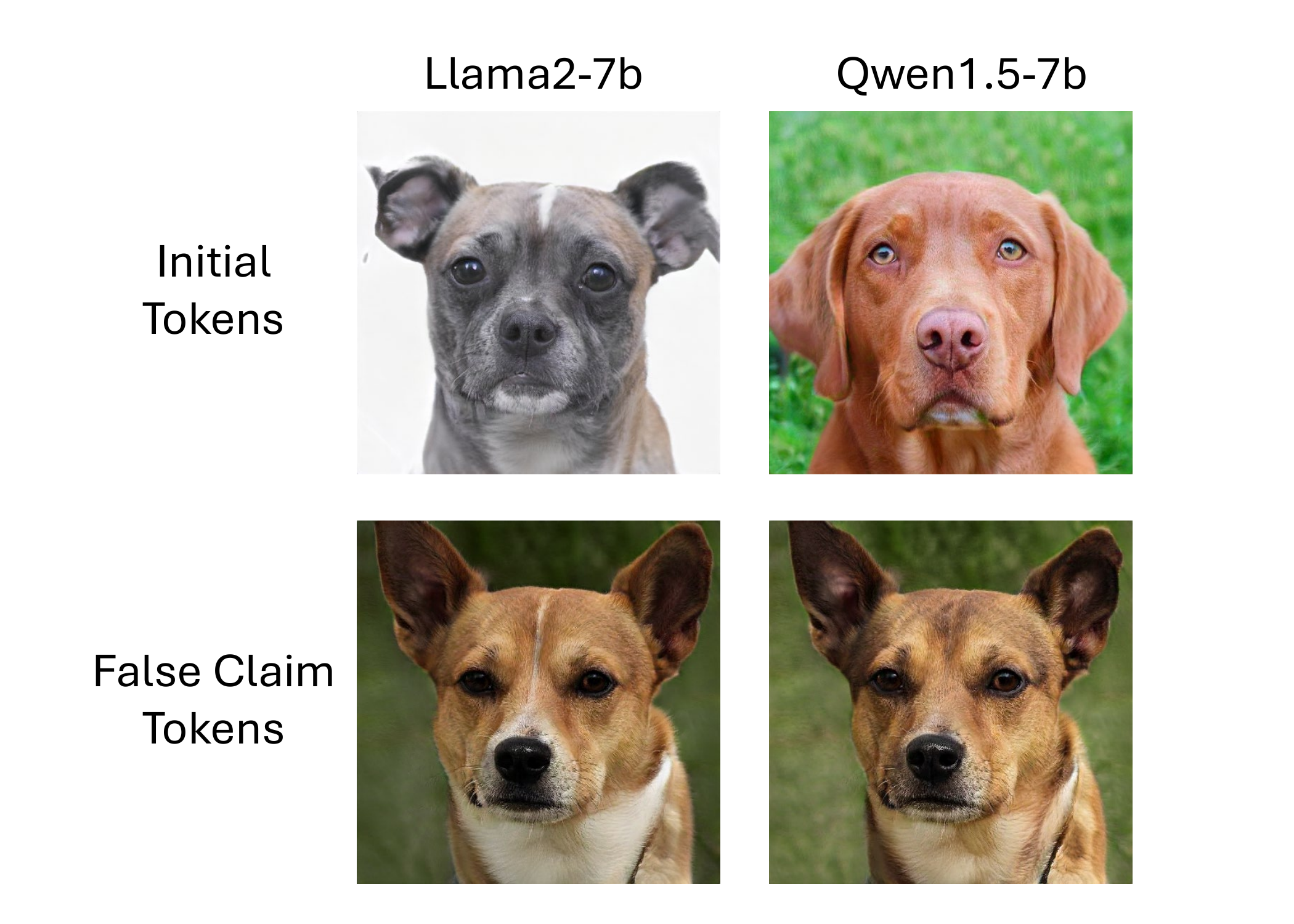}
    }
    \caption{False-claim attack on HuRef: both the feature vector and the human-readable fingerprint of the unrelated model (Qwen1.5-7B) become highly similar to that of the target model (Llama2-7B).}
    \label{fig:false_claim}
\end{figure*}

\rev{
\paragraph{False-claim attack on HuRef}
The same input-driven risk can be exploited more directly by optimizing the inputs to maximize fingerprint similarity.
We demonstrate this through a crafted false-claim attack against
\rev{HuRef-ICS}~\cite{zeng2024huref}, where an attacker crafts input tokens to make an unrelated model appear related to a model they hold. 
}

Fig. \ref{fig:huref_compair} shows the critical path in the similarity scoring mechanism of HuRef. Given an LLM, HuRef uses the input embeddings \(E\) together with the model weights to compute invariant terms \(I\), which serve as the model’s fingerprint. The cosine similarity of \(I\)  is then used to detect potential IP infringement. In addition, HuRef encodes \(I\) into a feature vector \(v\) and generates human-readable fingerprints (images), where visual similarity indicates related models. \rev{This dependence on the input provides an attack surface: the invariant terms of two independent models can be forced to converge via carefully crafted inputs, fabricating a false lineage without changing either model.}

% We envisioned an attack scenario where the malicious actor (a malicious owner or model stealer) aims to falsely claim the ownership of an unrelated model to his/her target or stolen model. To this end, the attacker carefully crafts input tokens so that, when the feature vector is computed following HuRef, the unrelated model yields high similarity to attacker’s model, leading to deceptive readable fingerprints. 

%Since the embedding step ($X \rightarrow E$) is not differentiable, we choose the genetic algorithm to implement the attack.
\rev{
To construct such inputs, we use a genetic algorithm to search over discrete token sequences.
}
Specifically, we treat a token as a gene and an input as an individual, and define the fitness as the similarity between the feature vectors computed by the two models given that input. In each generation, we apply random gene mutations (i.e., token substitutions) to the population of input individuals and select those with the highest fitness (i.e., highest feature vector similarity) for crossover to produce a new population. After several generations, the input with the highest fitness in the final population is selected as the false claim input. Such an attack only manipulates the input without altering the model itself.

% The above comparison process can be concluded as a function: $\text{ICS} = F(X)$. Then we can construct adversarial input tokens $\hat{X}$ to manipulate the output $\text{ICS}$.

In the experiment, we assume the attacker holds Llama2-7B and he/she wants to deliberately claim the ownership of an unrelated model Qwen1.5-7B. By feeding the two models with a carefully crafted input obtained as described above, their feature vector similarity  increased from 28\% to 95\% after 100 generations, leading to a false claim of model piracy. Moreover, the generated readable fingerprints shifted from being completely unrelated to appearing similar when using the false claim tokens, indicating that fingerprint similarity can be artificially induced.

\rev{Together, these results highlight a shared risk in input-dependent fingerprinting: the measured similarity may reflect the choice or optimization of inputs rather than true model provenance. By completely decoupling fingerprint extraction from input data, SELF inherently neutralizes this input-driven attack surface. While an attacker could theoretically bypass this by directly fine-tuning the model to forge the fingerprint of an unrelated target (weight-space false claim), our evaluation in Sec.~\ref{sec: robust_veri} confirms that SELF robustly resists such efforts. Therefore, SELF ensures reliable provenance verification even against highly capable white-box adversaries. }
% Therefore, eliminating the influence of inputs on the verification process is necessary for reliable model provenance verification.}

\rev{\subsection{Empirical Validation of Spectral Stability}}

\begin{table*}[t]
%\color{blue}
\revcolor
\centering
\footnotesize
\caption{$L_2$ norm differences of weight matrices ($W_Q$, $W_K$) and invariant matrices ($X_\sigma$, $X_\lambda$) between Llama-2-7B base and its fine-tuned variants (First 8 layers).}
\label{tab:weight_invariant_diff}
\begin{tabular}{ccccccccc}
\toprule
\multirow{2}{*}{\textbf{Layer}} & \multicolumn{4}{c}{\textbf{Llama-2-7B-chat}} & \multicolumn{4}{c}{\textbf{CodeLlama-7B}} \\
\cmidrule(lr){2-5} \cmidrule(lr){6-9}
& $\boldsymbol{\|\Delta W_Q\|_2}$ & $\boldsymbol{\|\Delta W_K\|_2}$ & $\boldsymbol{\|\Delta X_\sigma\|_2}$ & $\boldsymbol{\|\Delta X_\lambda\|_2}$ & $\boldsymbol{\|\Delta W_Q\|_2}$ & $\boldsymbol{\|\Delta W_K\|_2}$ & $\boldsymbol{\|\Delta X_\sigma\|_2}$ & $\boldsymbol{\|\Delta X_\lambda\|_2}$ \\
\midrule
0 & 0.6441 & 0.9481 & 14.9893 & 18.6290 & 10.2179 & 16.3403 & 308.7895 & 414.0640 \\
1 & 0.5998 & 1.0492 & 19.8387 & 27.8605 & 12.1084 & 15.3005 & 405.8361 & 526.0145 \\
2 & 0.4577 & 0.6067 & 16.3128 & 20.4678 & 7.0955  & 7.2867  & 408.6551 & 438.2098 \\
3 & 0.4623 & 0.5108 & 13.0404 & 17.4805 & 6.0730  & 6.3525  & 343.4302 & 387.0950 \\
4 & 0.4217 & 0.4619 & 14.5159 & 17.8721 & 6.5702  & 6.7162  & 370.9959 & 403.0934 \\
5 & 0.4830 & 0.5354 & 14.5439 & 18.9411 & 7.0580  & 7.6406  & 390.9983 & 444.8851 \\
6 & 0.4981 & 0.5204 & 13.3252 & 18.7204 & 9.7579  & 9.3975  & 344.3633 & 355.4242 \\
7 & 0.4899 & 0.5093 & 12.8714 & 18.3787 & 6.5182  & 6.9982  & 306.1046 & 343.1328 \\
\bottomrule
\end{tabular}%
\end{table*}

\begin{table*}[h]
\revcolor
\centering
\footnotesize
\caption{$L_1$ norm differences of raw spectra ($\tilde{\sigma}$, $\tilde{\lambda}$), normalized spectra ($\sigma$, $\lambda$), and layer-wise model fingerprints ($\mathcal{F}^i$) (First 8 layers).}
\label{tab:spectrum_fingerprint_l1}
\begin{tabular}{ccccccccccc}
\toprule
\multirow{2}{*}{\textbf{Layer}} & \multicolumn{5}{c}{\textbf{Llama-2-7B-chat}} & \multicolumn{5}{c}{\textbf{CodeLlama-7B}} \\
\cmidrule(lr){2-6} \cmidrule(lr){7-11}
& $\boldsymbol{\|\Delta \tilde{\sigma}\|_1}$ & $\boldsymbol{\|\Delta \tilde{\lambda}\|_1}$ & $\boldsymbol{\|\Delta \sigma\|_1}$ & $\boldsymbol{\|\Delta \lambda\|_1}$ & $\boldsymbol{\|\Delta \mathcal{F}^i\|_1}$ & $\boldsymbol{\|\Delta \tilde{\sigma}\|_1}$ & $\boldsymbol{\|\Delta \tilde{\lambda}\|_1}$ & $\boldsymbol{\|\Delta \sigma\|_1}$ & $\boldsymbol{\|\Delta \lambda\|_1}$ & $\boldsymbol{\|\Delta \mathcal{F}^i\|_1}$ \\
\midrule
0 & 29.6924 & 18.0723 & 0.0200 & 0.0978 & 0.1178 & 507.6064 & 257.8611 & 1.9152 & 1.3635 & 3.2786 \\
1 & 102.5995 & 55.8582 & 0.0254 & 0.1169 & 0.1423 & 436.0782 & 172.6528 & 1.0252 & 0.8056 & 1.8307 \\
2 & 84.8931 & 41.9358 & 0.0168 & 0.0894 & 0.1062 & 2196.6628 & 1199.6815 & 0.4681 & 0.9231 & 1.3912 \\
3 & 57.7201 & 30.8182 & 0.0159 & 0.0607 & 0.0765 & 2075.8862 & 1122.6079 & 0.7942 & 1.0659 & 1.8601 \\
4 & 61.4231 & 25.2875 & 0.0133 & 0.0907 & 0.1040 & 2110.1177 & 992.6726 & 0.4244 & 1.0701 & 1.4945 \\
5 & 59.3779 & 27.5637 & 0.0133 & 0.0925 & 0.1058 & 2426.4851 & 1184.1106 & 0.4847 & 0.7243 & 1.2090 \\
6 & 61.8070 & 31.8011 & 0.0149 & 0.0915 & 0.1064 & 2185.8218 & 1043.9443 & 0.2537 & 0.4562 & 0.7099 \\
7 & 57.0769 & 31.4931 & 0.0111 & 0.0522 & 0.0633 & 1823.5110 & 1136.3480 & 0.4854 & 0.4021 & 0.8875 \\
\bottomrule
\end{tabular}%
\end{table*}

\begin{table*}[h!]
%\color{blue}
\revcolor
\centering
\footnotesize
\caption{Condition number statistics of the top-256 eigenvalues for the invariant matrix $X_\lambda$ across Llama2-7B, Llama2-7B-chat, and CodeLlama-7B.}
\label{tab:ev_invariant_top256_eig_cond_combined}
\begin{tabular}{cccccccccc}
\toprule
\multirow{2}{*}{\textbf{Layer}} & \multicolumn{3}{c}{\textbf{Mean}} & \multicolumn{3}{c}{\textbf{Max}} & \multicolumn{3}{c}{\textbf{Min}} \\
\cmidrule(lr){2-4} \cmidrule(lr){5-7} \cmidrule(lr){8-10}
& \textbf{Llama2} & \textbf{Chat} & \textbf{Code} & \textbf{Llama2} & \textbf{Chat} & \textbf{Code} & \textbf{Llama2} & \textbf{Chat} & \textbf{Code} \\
\midrule
0 & 35.1765 & 31.1541 & 35.1168 & 170.0865 & 113.2376 & 103.7457 & 2.6866 & 2.7253 & 4.3320 \\
1 & 20.9765 & 21.7992 & 26.6926 & 89.5765  & 78.0695  & 228.4924 & 3.4169 & 3.2861 & 2.2162 \\
2 & 21.6719 & 26.3377 & 22.4314 & 62.7738  & 183.8452 & 170.6535 & 2.3612 & 2.3520 & 2.1549 \\
3 & 31.8259 & 20.7790 & 21.1383 & 930.7057 & 177.2809 & 82.8434  & 1.9210 & 1.9385 & 1.7110 \\
4 & 25.4155 & 22.7667 & 30.4696 & 217.4204 & 98.1534  & 155.5914 & 1.9837 & 2.0088 & 1.7797 \\
5 & 27.5151 & 25.1590 & 23.4014 & 173.8138 & 90.6868  & 95.6495  & 2.0320 & 2.0524 & 1.8756 \\
6 & 23.7864 & 24.1888 & 22.8176 & 118.8965 & 201.2819 & 152.0720 & 1.8212 & 1.8422 & 1.9557 \\
7 & 21.3125 & 19.9082 & 23.6527 & 73.5426  & 74.3570  & 351.4057 & 2.0859 & 2.0979 & 2.1604 \\
\bottomrule
\end{tabular}%
\end{table*}

% To further verify that the fingerprint deviation after downstream modifications can be bounded by Equation \ref{eq:weyl} and \ref{eq:fop}, we measured the actual perturbation magnitudes of the weight matrices and the variables used in the fingerprint computation for Llama-2-7B after fine-tuning it into Llama-2-7B-chat and CodeLlama-7B.
% Table \ref{tab:weight_invariant_diff} presents the $L_2$ norm differences of the attention weight matrices ($\Delta W_Q$, $\Delta W_K$) and their corresponding singular value / eigenvalue invariant matrices ($\Delta X_\sigma$, $\Delta X_\lambda$). 
% Furthermore, Table \ref{tab:spectrum_fingerprint_l1} details the $L_1$ norm variations of the raw and normalized values ($\tilde{\sigma}$, $\tilde{\lambda}$, $\sigma$, $\lambda$), alongside the overall shifts in the layer-wise model fingerprints ($\mathcal{F}^i$). 

% To validate the stability of the eigenvalues, Table \ref{tab:ev_invariant_top256_eig_cond_combined} reports the condition number statistics for the top-256 eigenvalues of the invariant matrix $X_\lambda$ across the three models. 
% The condition numbers indicate that the leading individual eigenvalues remain constrained, providing robust theoretical and empirical support for the reliability of the spectral shifts observed.
\rev{
To empirically assess the spectral stability characterized by Equations~\ref{eq:weyl} and~\ref{eq:fop}, we compared Llama-2-7B with its fine-tuned variants, Llama-2-7B-chat and CodeLlama-7B, across the first eight layers.
Table~\ref{tab:weight_invariant_diff} presents the $L_2$ norm differences of the attention weight matrices ($W_Q$, $W_K$) and their corresponding invariant matrices ($X_\sigma$, $X_\lambda$).
Table~\ref{tab:spectrum_fingerprint_l1} reports the $L_1$ norm differences of the raw spectra ($\tilde{\sigma}$, $\tilde{\lambda}$), normalized spectra ($\sigma$, $\lambda$), and layer-wise model fingerprints ($\mathcal{F}^i$).
Table~\ref{tab:ev_invariant_top256_eig_cond_combined} further summarizes the individual condition numbers of the top-256 eigenvalues of $X_\lambda$, which characterize the local sensitivity of the selected eigenvalues.

Across all evaluated layers and both fine-tuned variants, the observed raw singular-value differences remain strictly below the corresponding Weyl-based upper bounds.
Likewise, the observed raw eigenvalue differences remain well below the aggregate first-order sensitivity estimates obtained using the individual eigenvalue condition numbers of the base model and the measured invariant-matrix perturbations.
These results strictly align with the theoretical stability analysis, indicating that the observed spectral changes are substantially smaller than the conservative worst-case estimates.

Furthermore, $L_2$ normalization inherently compresses these bounded raw shifts into a highly compact decision space, allowing SELF to capture changes in invariant spectral shape rather than absolute scales. Consequently, the ultimate fingerprint distances for Llama-2-7B-chat and CodeLlama-7B sum to merely 0.8223 and 12.6612, respectively, both remaining safely below the verification threshold. Together, these measurements connect parameter perturbations to spectral and fingerprint deviations, providing robust theoretical and empirical support for SELF's stability under model modifications.
}

\section{Related Works}
\label{app:related_work}
Current model IP infringement detection methods mainly use watermarking or fingerprinting. Watermarking embeds identifiable features into the target models, while fingerprinting extracts unique identifiers without modifying the model, thus avoiding retraining and performance degradation.

\subsection{Watermarking} 
Watermarking approaches for LLMs primarily operate through two paradigms: model-centric watermarking and sampling-centric watermarking. 

Model-centric techniques focus on embedding watermarks by modifying the model’s architecture, parameters, or training process, typically involve fine-tuning or quantization strategies. For instance, \cite{xu2024hufu} leverages the permutation equivariance property of Transformers to embed watermarks by fine-tuning on strategically permuted data samples. Alternatively, \cite{li2024turningstrengthwatermarkwatermarking} introduces a knowledge injection approach where the watermark is carried through learned knowledge, while \cite{li2023watermarking} implements watermarking during the model quantization process. These approaches typically require additional retraining or fine-tuning costs. 

Sampling-centric methods embed watermarks by modifying the token sampling process or latent representations during text generation. These approaches do not alter the model’s parameters but instead tweak the decoding strategy or latent space.  \cite{zhang2024remark} presents REMARK-LLM, a framework that encodes both generated text and binary messages into latent space before transforming tokens into sparse distributions. \cite{dathathri2024scalable} develops Tournament sampling, a modified sampling algorithm for watermark embedding. The KGW method \cite{kirchenbauer2023watermark} employs a ``green list"  approach, preferentially selecting predetermined tokens during generation. However, such watermarking schemes remain susceptible to editing and spoofing attacks \cite{sadasivan2025aigeneratedtextreliablydetected}. To mitigate this issue, \cite{liu2024a} proposes a semantic invariant watermarking (SIR) method that generates watermarked tokens based on sentence-level embeddings. For sampling-centric approaches, additional post-processing may be required to verify watermarks. 

\subsection{Fingerprinting} 
Existing LLM fingerprinting methods encompass either behavioral or structural fingerprinting. 
Behavioral fingerprinting focuses on identifying LLMs by analyzing their output characteristics or input-response patterns. These methods leverage the unique behavior signatures that emerge from how models respond to specific inputs or adversarial prompts. \cite{gubri2024trap} employs adversarial suffixes to elicit model-specific responses, while \cite{iourovitski2024hideseekfingerprintinglarge} develops an evolutionary strategy using one LLM to identify distinctive features of others. \cite{pasquini2025llmmapfingerprintinglargelanguage} utilizes carefully designed queries to detect specific model versions. As these methods rely on observable input-output interactions, they are effective for black-box scenarios but remain vulnerable to attacks such as false claim attacks \cite{liu2024false}.

Structural fingerprinting methods analyze internal parameters or activation patterns. These methods exploit the unique structural properties of a model’s architecture or training dynamics. \cite{zeng2024huref} proposes HuRef, a human-readable fingerprint based on the stability of LLM parameter vector directions after pretraining convergence. While these vector directions demonstrate resilience to subsequent training, they remain vulnerable to direct weight manipulations such as permutation and linear mapping. To address this limitation, the authors develop three rearrangement-invariant fingerprint terms.
\cite{zhang2025reef} introduces REEF, a representation-based fingerprinting method employing Centered Kernel Alignment (CKA) to compute similarity score. \cite{wu2025gradientbasedmodelfingerprintingllm} presents TensorGuard, a framework extracting gradient-based signatures by analyzing their internal gradient responses across tensor layers to random input perturbations. However, these fingerprinting techniques also exhibit vulnerability to false claim attacks \cite{liu2024false} due to their inherent reliance on input engagement.

\section{Conclusion, \rev{Limitations, and Future Work}}
\label{sec:conclusion}

This paper presents SELF, a \rev{white-box structural} fingerprinting method that eliminates input dependency, preventing false claim attacks. By extracting fingerprints utilizing singular values and eigenvalues, we leverage their mathematical properties to guarantee uniqueness, scalability, and robustness. By evaluating the $L_1$ distance between the compact fingerprints, SELF facilitates \rev{highly reliable adversarial weight-lineage verification}. Our approach thus serves as a reliable tool for LLM IP forensics.

\rev{
A formal limitation of SELF is its inability to reliably identify sources that contribute knowledge solely through output distributions. Our experiments on FuseLLM-7B illustrate this limitation: SELF accurately identifies its initialization source, Llama2-7B, with a fingerprint distance of 0.4420, but fails to link it to the distribution-only sources, OpenLlama2-7B and MPT-7B, whose distances (56.4556 and 44.6037) exceed the decision threshold $\tau=24$.  These results are consistent with SELF's reliance on inherited weight structure. Consequently, distribution-only knowledge transfer lies outside SELF's intended scope, and a negative verification result does not rule out such behavioral infringement.

For broader model provenance verification and IP protection, future work may investigate combining structural fingerprinting with behavioral watermarking to capture complementary evidence of weight inheritance and distribution-based knowledge transfer. Developing such a framework requires determining whether behavioral watermarks survive fusion or distillation, and how orthogonal types of evidence can be integrated reliably. Formalizing such integrated systems and experimentally validating their robustness represent a valuable avenue for future research. 
}

\section*{Acknowledgment}

Google Gemini \cite{gemini} and ChatGPT \cite{chatgpt} are used to polish writing in sections from \ref{sec:intro} to \ref{sec:conclusion}.


% \section{References Section}
% You can use a bibliography generated by BibTeX as a .bbl file.
%  BibTeX documentation can be easily obtained at:
%  http://mirror.ctan.org/biblio/bibtex/contrib/doc/
%  The IEEEtran BibTeX style support page is:
%  http://www.michaelshell.org/tex/ieeetran/bibtex/
 
% \section{References}
% You can manually copy in the resultant .bbl file and set second argument of $\backslash${\tt{begin}} to the number of references
%  (used to reserve space for the reference number labels box).
 
\bibliography{IEEEabrv, references}
\bibliographystyle{IEEEtran}

 
%\vspace{5pt}

%\bf{If you include a photo:}
%\vspace{-33pt}
\begin{IEEEbiography}[{\includegraphics[width=1in,height=1.25in,clip,keepaspectratio]{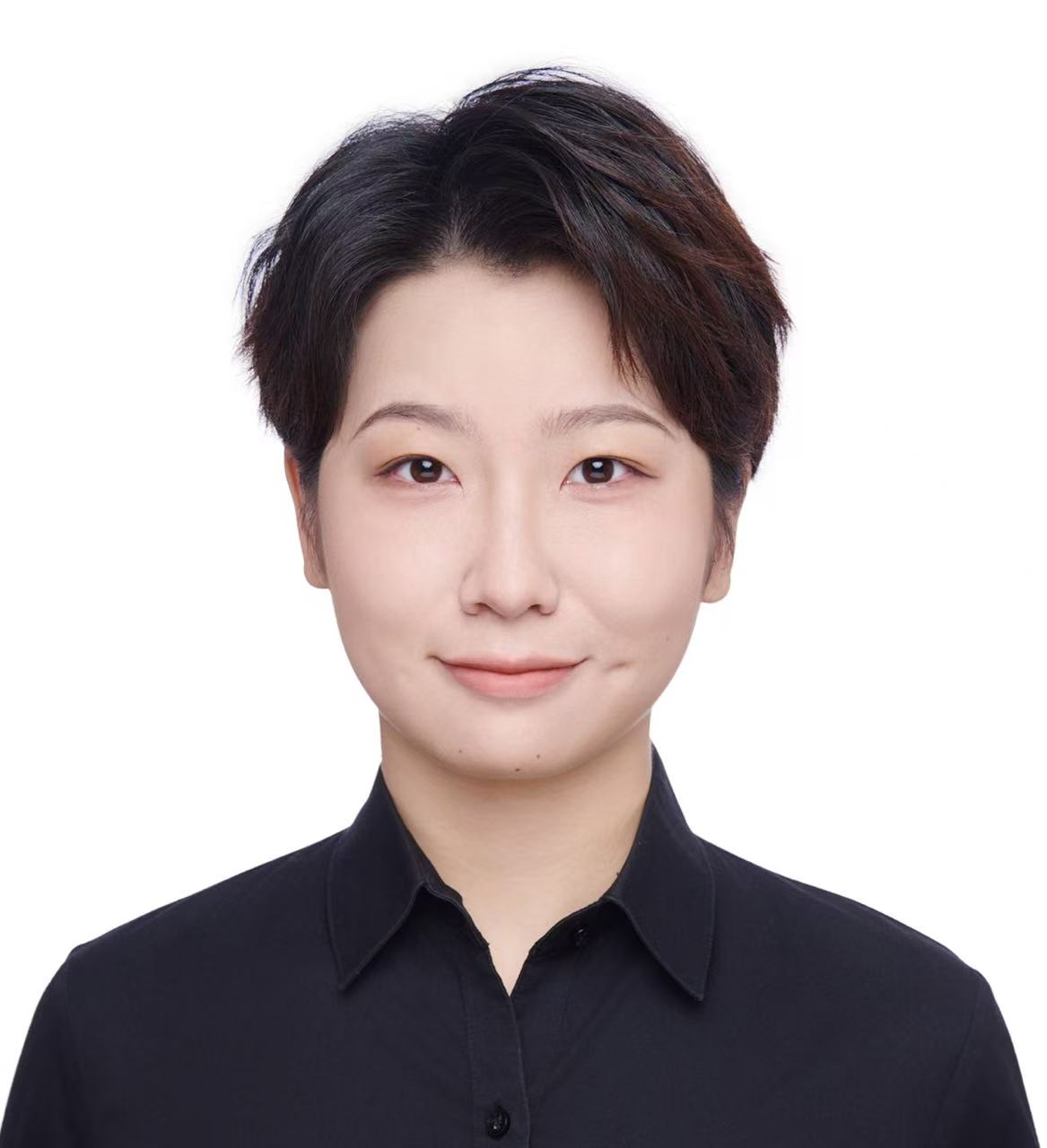}}]{Hanxiu Zhang}
received the B.E. degree from Northeastern University, Shenyang, China, and the M.E. degree from East China Normal University, Shanghai, China. She is currently working toward the PhD degree with the Chinese University of Hong Kong, Shenzhen. Her research interests include AI security.
\end{IEEEbiography}

%\vspace{5pt}

\begin{IEEEbiography}[{\includegraphics[width=1in,height=1.25in,clip,keepaspectratio]{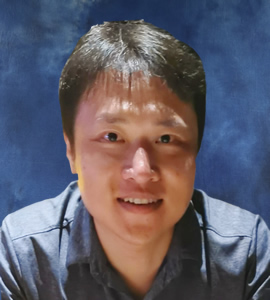}}]{Qiang Xu}
(Senior Member, IEEE) received the B.E. and M.E. degrees from Beijing University, Beijing, China, in 1997 and 2000, respectively, and the Ph.D. degree from McMaster University, Hamilton, ON, Canada, in 2005. He is a Professor of Computer Science and Engineering with The Chinese University of Hong Kong, Hong Kong. His research interests include electronic design automation, trusted computing, and representation learning. He is currently serving as an Associate Editor for IEEE Transactions on Computer-Aided Design of Integrated Circuits and Systems and Integration, the VLSI Journal.
\end{IEEEbiography}

%\vspace{5pt}

\begin{IEEEbiography}[{\includegraphics[width=1in,height=1.25in,clip,keepaspectratio]{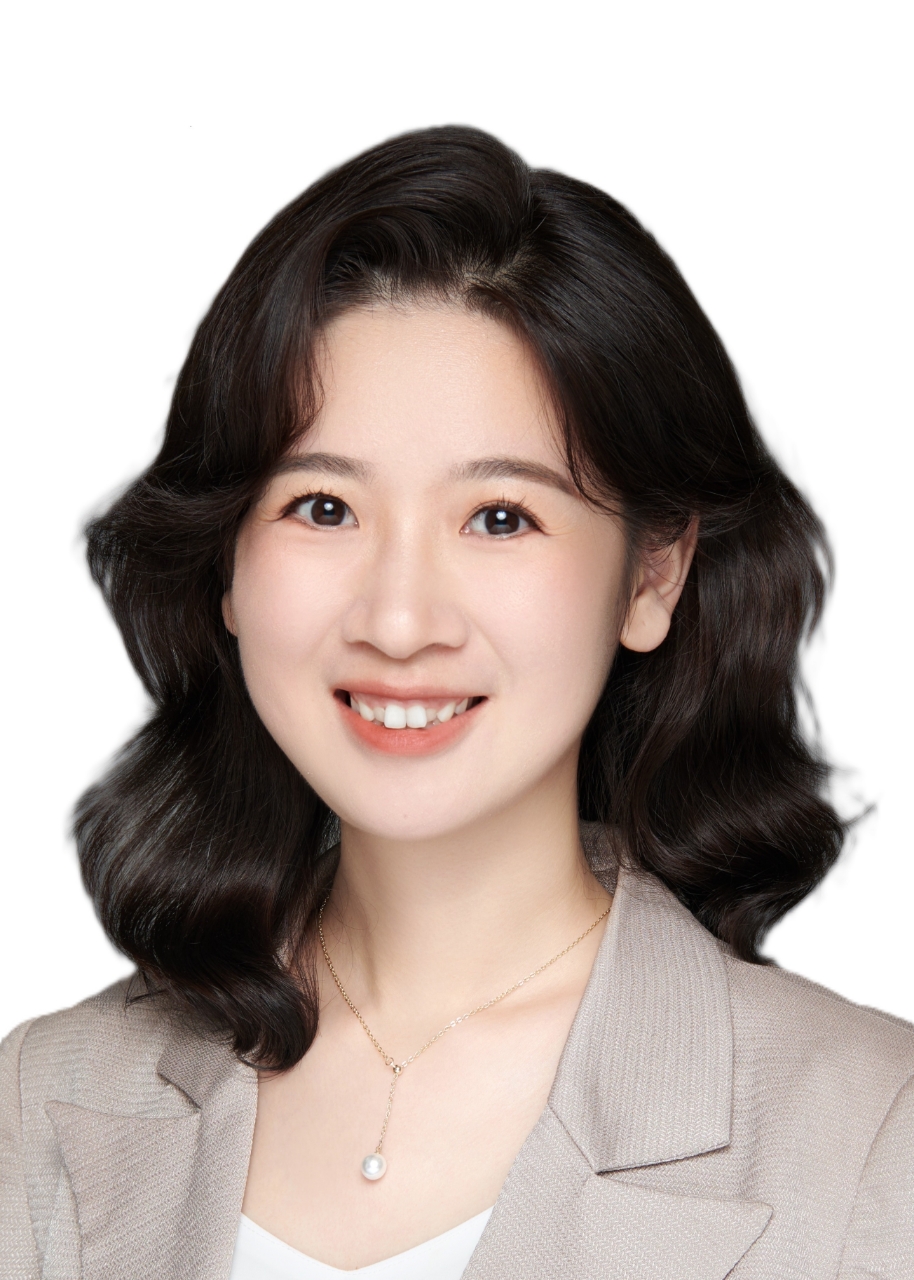}}]{Yue Zheng}
(Member, IEEE) received the BEng degree from Shanghai University (SHU, China), and the PhD degree from Nanyang Technological University (NTU, Singapore), in 2015 and 2020, respectively. She is currently an Assistant Professor with the Chinese University of Hong Kong, Shenzhen. Her areas of research include hardware security, security protocol design and trustworthy AI. She is currently serving as an Associate Editor for IEEE Transactions on Information Forensics and Security, and EURASIP Journal on Information Security. 
\end{IEEEbiography}


\vfill
\end{document}